\documentclass[a4paper,11pt,notitlepage]{article}
\pdfoutput=1 

\usepackage{jcappub} 

\usepackage{tikz}
\usepackage{graphicx}
\usepackage{dcolumn}
\usepackage{bm}

\usepackage{tabularx}

\usepackage{physics}
\usepackage{comment}
\usepackage{multirow}
\usepackage{subfigure}
\usepackage{xcolor}

\newcolumntype{Y}{>{\centering\arraybackslash}X}
\newcolumntype{Z}[1]{>{\centering\arraybackslash}m{#1}}
\usepackage{cellspace}
\graphicspath{{figures/}}

\title{Asteroseismological constraints on---and hints of---dark matter interactions}
\author[a,b]{Stephanie Beram}
\author[a,b,c]{Aaron C. Vincent}
\affiliation[a]{
Department of Physics, Engineering Physics and Astronomy,\\
Queen’s University, Kingston ON K7L 3N6, Canada
}
\affiliation[b]{%
Arthur B. McDonald Canadian Astroparticle Physics Research Institute, Kingston ON K7L 3N6, Canada
}
\affiliation[c]{Perimeter Institute for Theoretical Physics, Waterloo ON N2L 2Y5, Canada}
\emailAdd{stephanie.beram@queensu.ca}
\emailAdd{aaron.vincent@queensu.ca}
\date{\today}

\abstract{If dark matter interacts with nuclei or electrons, then elastic collisions with constituents of stars will cause some of the galactic dark matter to fall below the escape velocity and become gravitationally bound. For asymmetric dark matter (which does not self-annihilate), the large accumulated population of dark matter can act as an additional source of heat transport, altering stellar structure and evolution. These effects can be probed by the use of asteroseismology. Here, we demonstrate this effect via numerical simulations. We use Monte Carlo-calibrated heat transport calculations, with a focus on the erasure of the convective core in stars that are slightly more massive than the Sun. We find limits on spin-dependent dark matter-nucleon and dark matter-electron interactions using asteroseismological data from a nearby sub-giant star. More tantalizingly, we find a $\gtrsim 4 \sigma$ preference for dark matter-electron interactions for dark matter masses $\lesssim 3.5$ GeV and cross sections $\sigma_{\chi-e} \sim 10^{-34.5}$ cm$^2$, albeit in strong tension with limits from Earth-based direct detection experiments.
}

\begin{document}
\maketitle

\section{Introduction}
Observations indicate that dark matter (DM) constitutes nearly 85\% of the Universe’s matter density \cite{OriginalDM} though its particle nature remains unknown. Numerous  models offer DM particle candidates via extensions to the Standard Model (SM). One of the most widely studied solutions involves the introduction of a new, broadly-defined ``Weak-scale'' particle, i.e. with a mass in the GeV-TeV range, with interactions suppressed by a small coupling, massive mediator, or Lorentz structure.

One of the most promising and mature methods for WIMP searches remains direct detection \cite{goodman1985detectability} (DD), the search for DM elastic scattering-induced nuclear or electronic recoils in highly sensitive, shielded detectors. Currently, constraints on spin-independent (SI) DM-nucleus scattering cross sections are now well below $10^{-47.5}$ cm$^2$ for a DM mass $m_\chi \simeq 30$ GeV  and $10^{-41}$ cm$^2$ for spin-dependent (SD) interactions \cite{LZ:2024zvo}. 
 
If dark matter can scatter with detector material in the lab, it can necessarily do so in stars and planets. Indeed, there has been significant exploration of DM detection prospects through its influence on astrophysical objects such as black holes \cite{DMeffectBH, eda2013new, kazunari2015gravitational, boucenna2018novel, macedo2013into, yue2018gravitational}, neutron stars \cite{goldman1989weakly, DMeffectBHNS, bramante2014detecting, de2010neutron, mcdermott2012constraints, kouvaris2011constraining, kouvaris2014growth}, main sequence  \cite{renzini1987effects, lopes2019asymmetric, Iocco:2012wk, taoso2010effect, zentner2009high,Vincent:2013lua,Vincent:2014jia,vincent2015generalised, raghuveer2017dark,raen2021effects,John:2023knt,Banks:2021sba,Croon:2024waz,John:2024thz} and post-main sequence stars \cite{Lopes:2021jcy,Hong:2024ozz}, planets \cite{Leane:2020wob,Croon:2023bmu,Blanco:2023qgi,Benito:2024yki,Blanco:2024lqw,Leane:2024bvh,Acevedo:2024zkg,Robles:2024tdh}, and many other objects. 

For DM to affect a star, it must be captured via a scattering event that leaves it with a speed lower than the local escape velocity from the star \cite{Press:1985ug,Gould:1987ju, Gould:1987ir}. Once captured, the DM rapidly thermalises within the star, falling into an equilibrium distribution near the core. There are two scenarios which can arise from a captured DM population: 
1) if the dark matter can self-annihilate to SM particles \cite{Hagelin:1986gv, Srednicki:1986vj, Griest:1986yu, Silk:1985ax, Krauss:1985aaa, Freese:1985qw, Gaisser:1986ha}, it may lead to extra heating  and emission of neutrinos (detectable at Earth, in the case of the Sun \cite{barger2002indirect, desai2004search, Super-Kamiokande:2004pou,Rott:2012qb,Bernal:2012qh,Rott:2012qb,Bernal:2012qh,IceCube:2012ugg,IceCube:2012fvn, IceCube:2016yoy}), and 2) if the dark matter cannot self-annihilate, i.e. is \textit{asymmetric} (A)DM  \cite{zurek2014asymmetric, Petraki:2013wwa}, then its long mean free path in the stellar interior means that it acts as an efficient heat conductor \cite{Spergel:1984re, gould1990thermal, gould1990cosmion}, potentially altering the structure of the star itself through its equation of state. 

 Our focus is on the latter possibility. This idea was first suggested in the context of the solar neutrino problem \cite{solarNeutrinoProblem1, Spergel:1984re}. Though this was resolved by the confirmation of neutrino oscillation, the physical prediction remains: any asymmetric dark matter particle with a nonzero cross section with standard model particles will, to some degree, scatter and be captured by stellar nuclei or electrons and lead to a change in stellar structure. 
 
Indeed, heat transport by ADM has more recently been examined as a possible solution \cite{frandsen2010asymmetric,taoso2010effect,Cumberbatch:2010hh, lopes2014helioseismology, vincent2015generalised,Vincent:2014jia}  to the solar composition problem, a discrepancy between the standard solar model (SSM) predictions using measured elemental abundances, and helioseismology \cite{prieto2001forbidden, AllendePrieto:2002gnt, Melendez:2008jj, Scott:2008vt, Bahcall:2004yr, Basu:2004zg, Bahcall:2005va, Bi:2011sy}.

ADM can of course accumulate and impact the properties of other main-sequence stars. Energy transport by DM is expected to reduce the core temperature, and increase the core density and pressure, leading in turn to changes in the stellar structure and evolution. 
Stellar simulations which include the impact of ADM have shown that it can affect the star's main sequence lifetime and trajectory on the Hertzsprung-Russel (HR) diagram  \cite{lopes2019asymmetric, raen2021effects}, and asteroseismic observables \cite{Casanellas:2012jp, Martins:2017fji, Rato:2021tfc}. 

One promising signal of ADM originates in stars predicted to develop a convective core. 
Stars with masses $\lesssim 1.2 \ M_\odot$ are powered primarily by the proton-proton (pp), which has a relatively low dependence on temperature, allowing for a shallow gradient and hydrostatic equilibrium throughout the core, and heat transport via radiation alone.
In contrast, stars with masses $\gtrsim 1.2\ M_\odot$ rely on the CNO cycle, which results in a steep temperature gradient and a breakdown of local hydrostatic equilibrium and therefore convective motion in the core.
The heat transported by ADM can smooth out the temperature gradient by taking energy from the core and depositing it in the outer layers, which can lead to the re-establishment of hydrostatic equilibrium and consequently the erasure of the convective core. This was first suggested by Renzini \cite{renzini1987effects} to show convective core suppression in horizontal branch stars, which resulted in large modification in asymptotic giant branch lifetimes. The presence---or absence---of a convective core is a well-understood target for asteroseismology, as the sharp change in sound speed versus radius leads to partial reflection of travelling waves, and manifests as a distinct change eigenfrequencies observed in stellar oscillations. 

Recent years have seen space-based missions such as CoRot \cite{baglin2008corot} and Kepler \cite{borucki2010kepler} detecting oscillations in thousands of stars, providing unprecedented insights into their interiors. The data not only allow for more accurate measurements of a star's mass and radius, but also  better modelling of the stellar core and detection of a convective region. Such measurements were used in Ref.~\cite{Casanellas:2015uga}, who modeled ADM heat transport in a star imaged by the Kepler spacecraft.

To assess the impact of DM on stars, it is crucial to correctly model its capture rate and heat transport over the star's evolutionary history. Recent work \cite{Banks:2021sba,Banks:2024eag} has shown that the two commonly-used formalisms lead to an incorrect parametrisation of either the magnitude or radial profile of transported heat. To solve the Boltzmann equation responsible for heat transport by a weakly interacting species, one may assume an isothermal DM distribution for very long mean free paths $l_\chi$ relative to the DM scale radius $r_\chi$ \cite{Spergel:1984re} (i.e. large Knudsen number $K \equiv l_\chi/r_\chi$), leading to a relatively simple expression for heat transport. Though easier to compute, this has been known to be off by a factor of $\sim$2 \cite{gould1990cosmion,Banks:2021sba}, even when evaluated in the isothermal regime. At the opposite end, the LTE approach \cite{gould1990thermal} works well when $K \ll 1$ at intermediate radii. However, the method is plagued by numerical instabilities, it requires a correction near the core of a star, and the additional ``Knudsen correction'' used in more recent literature to cover smaller cross sections does not lead to a robust result \cite{Banks:2021sba}. In Refs. \cite{Banks:2021sba,Banks:2024eag} we demonstrated by direct Monte Carlo simulation that a simple rescaling of the  isothermal solution actually reproduces exact solutions across a vast range of interaction strengths and scalings (i.e. not just constant, but also $\sigma \propto v^{2n}, q^{2n}$, powers of the relative velocity and exchanged momentum, respectively), dark matter masses, and stellar models including main sequence stars, brown dwarfs, and planets. As these are new results, no observable predictions have yet been made with them.

For DM to be captured and transport energy in the star, it needs to interact with a target species inside the star. Most works considering ADM focus on DM-nucleon interactions. In this case, DM capture and energy transport are most efficient in the $\sim$ few GeV mass range, as the DM is best kinematically matched for scattering with hydrogen and helium. However, it is also possible that the DM has no tree-level coupling to quarks and scattering off electrons is the dominant interaction channel, as in e.g. \cite{Ibarra:2009bm,Essig:2011nj,Foot:2014xwa}\footnote{Even if DM couples only to quarks, loop-level interactions with electrons will almost certainly be present. These can even dominate for low DM masses, thanks to kinematics \cite{Diamond:2023fsm}.}. In this work, we will examine the impact of DM capture and energy transport in main-sequence stars due to interactions with either electrons and nucleons.

We therefore return to asteroseismological modeling of nearby stars, but with the above-mentioned improved modelling of the DM heat transport. We additionally perform self-consistent calibration of the target stellar model, which turns out to be crucial in correctly accounting for nuisance parameters. We additionally compare DM-nucleon versus DM-electron scattering. Altogether, this will will lead not only to stronger constraints, but also a hint of possible new physics in a nearby star. 

This article is organized as follows: in Sec.~\ref{sec:captRate}, we describe the capture and energy transport formalisms. Sec.~\ref{section:other_star} discusses our implementation in the \texttt{MESA} software suite, and results of heat transport by DM scattering respectively on nuclei and electrons in stars near the threshold to develop a convective core. Sec.~\ref{sec:realstar} then focuses on the impact of DM on the subgiant (SG) KIC 8228742, using observational data to constraint DM models. We conclude in Sec.~\ref{sec:conclusion}. Throughout this work, we focus on constant DM-SM cross sections. However, we provide in Appendix \ref{sec:appendix} some more detailed derivations and expressions for non-constant cross sections, as they correct some previous errors in the literature and may be helpful for future work. 

\section{Dark Matter Capture and heat transport in Stars}\label{sec:captRate}

In this section, we summarize the capture and heat transport formalisms used in our stellar simulations. 

We call the DM-nucleon or DM-electron differential cross section $d\sigma/d\cos \theta \equiv \sigma_0$. For DM-electron interactions, we treat the plasma as fully ionized, so the single-particle interaction cross section is simply $\sigma_0$, and the electrons are free, and in local thermal equilibrium. For DM-nucleus interactions, we focus on spin-dependent (SD) interactions, and only include scattering with hydrogen. Although some heavier elements with nonzero nuclear spin exist inside main-sequence stars, they do not benefit from a coherent enhancement so always remain subdominant to scattering with H due to their low abundances (see e.g. Fig. 1 of Ref.~\cite{Catena:2015uha}). 

Spin-independent scattering can of course also lead to capture, and can be especially beneficial in hydrogen-depleted regions such as the degenerate cores of red giants \cite{Lopes:2021jcy}. Even in such cases, stellar constraints have difficulty competing with modern earth-based experiments \cite{Hong:2024ozz}.

As already mentioned, we will focus on constant DM-target cross sections, but will present some key results for cross sections proportional to an even power $2n$ of the momentum transfer $q^{2n}$ or relative velocity $v^{2n}$, which can arise generically in effective theories of dark matter \cite{Fan:2010gt}. Mathematical results for these cases are derived in Appendix \ref{sec:appendix}.

\subsection{Dark matter capture rate}\label{DM_Capture_Rate}

If a DM particle scatters off a target, $T$, inside a star, it can lose enough kinetic energy for its velocity to fall below the local escape velocity $v_e(r) = \sqrt{-2 \phi(r)}$ (where  $\phi (r)$ is the gravitational potential) and consequently become gravitationally bound to the star. The DM capture rate, $C$, is given by
\begin{equation}\label{equation:captureRateRaw}
C = \int_0^{R_\star} 4 \pi r^2 dr \int_0^{\infty} d u_\chi \frac{\rho_\chi}{m_\chi} \frac{f_{v_\star}(u_\chi)}{u_\chi} w(r) \int_0^{v_e(r)} R^-_T (w \rightarrow v) dv, 
\end{equation} 
where $R_\star$ is the star's radius, $u_\chi$ is the DM velocity at infinity, $\rho_\chi$ is the DM local density, $m_\chi$ is the DM mass, $v_\star$ is the star's velocity, and $f_{v_\star}(u_\chi)$ is the halo velocity distribution in the star's reference frame moving at $v_\star$ relative to the Galactic rest frame given by 
\begin{equation}
    f_{v_\star}(u_\chi) = \frac{1}{2} \int_{-1}^1 f_{\text{gal}} \left( \sqrt{u_\chi^2 + v_\star^2 + 2 u_\chi v_\star \cos \theta_\star}\right) d \cos \theta_\star,
\end{equation}
where $\cos \theta_\star$ is the angle between the DM and star's velocities, and $f_{\text{gal}}$ is the DM velocity distribution in the Galaxy's reference frame which is assumed to be a Maxwell-Boltzmann distribution with velocity dispersion $v_d = \sqrt{3/2} \ v_\star$. The integral yields
\begin{equation}
f_{v_\star}(u_\chi) = \left( \frac{3}{2} \right)^{3/2} \frac{4 \rho_\chi u_\chi^2}{\sqrt{\pi} m_\chi v_d^3} \exp \left(  - \frac{3(v_\star^2+u_\chi^2)}{2\: v_d^2} \right) \frac{\sinh \left( 2 u_\chi v_d / v_d^2 \right)}{3 u_\chi v_\star / v_d^2}. 
\end{equation}
The factor $R^-_T (w \rightarrow v)$ is the differential scatting rate, representing the rate at which a DM particle with initial velocity $w$ scatters off a target $T$ to a lower final velocity $v$. This rate also hinges on the type of interaction, DM interaction strength, and the target's density and mass, and is given by:

\begin{equation}\label{equation:diffScatteringRateRaw}
R_T^- (w \rightarrow v) = \int n_T(r) \frac{d \sigma}{d v} |\boldsymbol{w}-\boldsymbol{u}| f_T(u, r) d^3 \boldsymbol{u}, 
\end{equation}
where $n_T(r)$ is the number density of the target (electron or nucleon) population, $d\sigma / dv$ is the differential scattering cross section, and $ f_T(u, r)$ the target's distribution function given by a Maxwell-Boltzmann distribution.  We can expand the 3D integral and insert in the target velocity distribution explicitly to obtain







\begin{equation}\label{RRate_raw}
R_T^- (w \rightarrow v) = \frac{2}{\sqrt{\pi}} \frac{n_T(r)}{u_T^3(r)} \int_0^\infty du\: u^2 \int_{-1}^1 d \cos \theta \frac{d \sigma}{dv} |\boldsymbol{w}-\boldsymbol{u}| e^{-u^2/ u_T^2},
\end{equation}
where $u_T = \sqrt{2 T_\star(r) / m_T}$ is the thermal speed of the target particles with mass $m_T$ and temperature $T_\star(r)$. This integral is then evaluated by moving from the lab frame to the center of mass frame and expressing the integral as a function of the velocity  of the center of mass and the velocity of the incoming DM particle in the center-of-mass frame as described in Appendix A of \cite{raghuveer2017dark}. For constant cross sections, as well as for and $q$ or $v$ dependent interactions with positive powers (see Appendix \ref{sec:appendix}), the integral can be evaluated analytically. We use the notation  $R_{T, n_q, n_v}^- (w \rightarrow v)$ to denote the differential scattering rate for a particular interaction. For a constant (velocity-independent and isotropic) cross section, this rate is given by \cite{Gould:1987ju}
\begin{equation}
    R_{T, 0, 0}^- (w \rightarrow v) = \frac{4}{\sqrt{\pi}} \frac{\mu_+^2}{\mu} \frac{v}{w} n_T(r) \sigma_0 \left[ \chi(-\alpha_-, \alpha_+) + \chi(-\beta_-, \beta_+) e^{\mu(w^2-v^2)/u_T^2(r)}\right],
\end{equation}
where we have defined the usual $\mu = m_\chi/m_T$, $\mu_{\pm} = (\mu \pm 1)/2$, and 

\begin{equation}
    \chi(x, y) \equiv \int_x^y e^{-y^2} dy = \frac{\sqrt{\pi}}{2} \left[\text{Erf}(y) - \text{Erf}(x) \right],
\end{equation}

\begin{equation}
    \alpha_\pm \equiv \frac{\mu_+ v \pm \mu_- w}{u_T(r)},
\end{equation}

\begin{equation}
    \beta_\pm \equiv \frac{\mu_- v \pm \mu_+ w}{u_T(r)}.
\end{equation}
We can then evaluate the total capture rate \eqref{equation:captureRateRaw} by performing the integrals over the DM velocity $u_\chi$ and the radius $r$. To perform the radial integral, a stellar model is needed to specify the radial profiles of the star's temperature $T(r)$, gravitational potential $\phi(r)$, and target number density $n_T(r)$.

\begin{figure}
    \centering
    \includegraphics[width=0.5\linewidth]{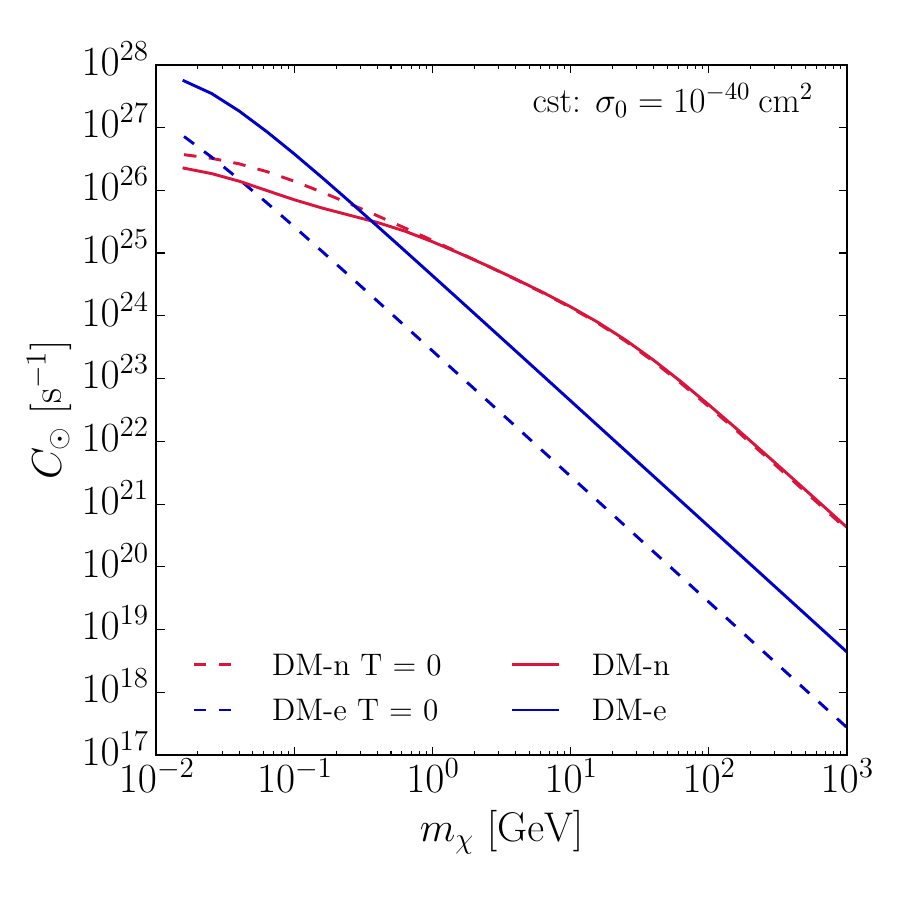}
    \caption{Capture rate of dark matter interacting with nucleons (red) and electrons (blue) inside the Sun. Computations including thermal motion of the target are shown as solid lines; the zero-temperature approximation is shown with dashed lines. }
    \label{fig:capratev0q0}
\end{figure}

Previous work has shown that for DM-nucleon interactions, the effect of the target species' thermal motion on the capture rate is negligible when considering DM masses which are comparable or more massive than the target species \cite{raghuveer2017dark}. This allows setting their temperature to zero, as detailed in Appendix C of \cite{Busoni:2017mhe}. 
This approximation simplifies the differential scattering rate calculation significantly, which can be beneficial for numerical evaluations given that it needs to be integrated over the final (post-scattering) DM velocities at each radial shell of the star. 

These thermal effects are proportional to the thermal speed of the target species $u_T = \sqrt{2 \ T_\star/m_T}$ (with $T$ expressed in natural units, i.e. $k_B = 1$) and are therefore far more important for lighter target species. This will mean that thermal motion cannot be neglected when computing capture via interaction with electrons. 

We show the resulting capture rates in the Sun\footnote{We use a standard solar model from \href{https://www.ice.csic.es/personal/aldos/Solar_Models.html}{https://www.ice.csic.es/personal/aldos/Solar\_Models.html} \cite{vinyoles2017new} } in Fig.~\ref{fig:capratev0q0}. With (solid) and without (dashed) accounting for thermal motion of the target. This clearly matters for capture on electrons, but not nucleons. It is further clear that the low target mass means that electrons have a much harder time capturing dark matter with masses above $\sim 1$ GeV. Equivalent plots for non-constant cross sections can be found in Appendix \ref{sec:appendix}.

\subsection{Evaporation}
 Evaporation occurs when DM particles are up-scattered to velocities exceeding the local escape velocity of the star, causing them to leave the star and thereby reducing the overall number of DM particles within it. This  becomes relevant for lower DM masses because lighter DM particles are more likely to achieve the necessary speed to exceed the stellar core’s escape velocity. 

Detailed evaporation mass calculations have been performed in Ref. \cite{Busoni:2017mhe} for DM-nucleon interactions, and in Ref~\cite{raghuveer2017dark} for both interactions with nucleons and electrons. For nucleons, evaporation occurs below DM of $\sim 3-4.5$ GeV. The evaporation mass for electrons is likely similar, though the authors of Ref.~\cite{raghuveer2017dark} point out that a small cutoff in the DM velocity distribution (due to prior evaporation) can lead to sub-GeV evaporation masses. However, this would require thermal equilibration timescales to be much slower than the evaporation timescale, which seems unlikely.

\subsection{Geometric Limit}\label{subsection:geomLimit}

The total effective cross-section of a star for collisions with DM particles cannot surpass the geometric limit
\begin{equation}
\sigma_{\text{sat}} = \pi R_\star^2(t),
\end{equation}
which is the cross section where the star is optically thick to DM. 
This results in an upper limit on the capture rate,
\begin{equation}\label{equation:satCrossSection}
C_{\text{geom}} = \frac{1}{3} \pi \frac{\rho_\chi}{m_\chi} R_\star^2 \left( e^{- \frac{3}{2}\frac{u_\star^2}{u_0^2}} \sqrt{\frac{6}{\pi}} u_0 + \frac{6\: G_N M_\star + R_\star \left( u_0^2 + 3 \: u_\star^2 \right)}{R_\star u_\star} \text{Erf} \left[ \sqrt{\frac{3}{2}} \frac{u_\star}{u_0} \right] \right),
\end{equation}
where $G_N$ is Newton's gravitational constant. Therefore, we model the capture rate as the lesser between the rates calculated using Eqs. \eqref{equation:captureRateRaw} and \eqref{equation:satCrossSection}. 

    \subsection{Dark matter energy transport}\label{section:energyTransport}

Once DM is captured, it thermalises and accumulates near the centre of the star. The DM population can then act as an alternative channel for macroscopic energy transport by absorbing energy through elastic interactions with a target species in the star's interior and then transporting this energy and depositing it in the cooler outer layers of the star. This energy transport can result in modifications to the star's structure and evolution by altering its temperature, density, sound speed, and pressure radial profiles. 

This heat transport depends on the details of the DM-target interactions, as well as the star's properties such as its mass and radius, its gravitational potential $\phi(r)$, temperature $T(r)$, density $\rho(r)$, and element abundances. The heat transport is governed by a Boltzmann Collision Equation (BCE) 

\begin{equation}
D F = l_\chi^{-1} C F,
\label{eq:BCE}
\end{equation}
where $F = F(\boldsymbol{r},\boldsymbol{v},t)$ is the DM phase space distribution, $D$ is the differential Liouville operator, and $C$ is the collision operator which accounts for the scattering events. The length $l_\chi$ is the typical inter-scattering distance:
\begin{equation}\label{equation:mfp}
l_\chi = \frac{1}{n_T(r) \langle \sigma (\boldsymbol{v}) \rangle},
\end{equation}
where $n_T$ is the target species' number density, and $\langle \sigma (\boldsymbol{v}) \rangle$ is the thermally averaged (total) cross section. For a constant cross section, this is simply $\langle \sigma (\boldsymbol{v}) \rangle = 2\sigma_0$. 

The thermal behaviour of the captured DM population is described by the Knudsen number
\begin{equation}\label{equation:KnudsenNumber}
K = \frac{l_\chi(0)}{r_\chi}, 
\end{equation}

which is the ratio between the mean inter-scattering distance at the centre of the star and a DM scale height $r_\chi$, which can be derived from the Virial theorem, given by 

\begin{equation}
r_\chi = \sqrt{\frac{2 T_c}{2 \pi G_N \rho_c m_\chi}}, 
\end{equation}
where, $T_c$ is the central temperature, $\rho_c$ is the central density, $G_N$ is Newton's gravitational constant, and $m_\chi$ is the DM mass. 

Eq. \eqref{eq:BCE} is a 7-dimensional integro-differential equation. Even though it is simplified by considering the steady state ($\partial F/\partial t$ = 0), and by rotational symmetry, it remains analytically intractable. It can, however, be approximately solved in two limiting regimes: 
\begin{itemize}
\item[•] $K \gg 1$: the large mean-free-path regime which was explored by Spergel $\&$ Press (SP) \cite{Spergel:1984re}. In this regime, the DM population is described as an \textit{isothermal} heat bath with an single temperature $T_\chi$. The energy transport occurs due to the temperature difference between the DM and local plasma populations. This regime would typically have a smaller interaction cross section and allows DM to travel large distances within the star. Heat transport is proportional to the cross section. 
\item[•] $K \ll 1$: this is a local-thermal-equilibrium (LTE) regime which was explored by Gould $\&$ and Raffelt (GR) \cite{Gould:1987ju}. In this regime, the DM population is said to be in local equilibrium with its surroundings meaning that the DM temperature tracks the temperature of the stellar plasma $T_\chi(r) = T_\star(r)$. This regime corresponds to larger interaction cross sections leading to a high collisional frequency. Heat transport is inversely proportional to the cross section. 
\end{itemize} 

In the SP regime, the transport is enhanced by the longer inter-scattering distance which allows DM to transport energy to larger radii. However, the small cross sections lead to a reduction in collision efficiency and therefore the amount of energy DM is transports is suppressed. 
In the GR regime, the strong interaction cross section leads to high collisional frequency which enhances energy absorption and deposition by DM, however DM is stuck within a small radial shell and cannot transport the heat very far.

Transport is maximal for intermediate cross sections where $K \approx 1/3$. However in this regime, neither of the approximation schemes are valid, and therefore the BCE cannot be solved analytically. This has led authors to correct one formalism or another, to account for the behaviour on traversing the Knudsen peak. In Refs. \cite{Banks:2021sba,Banks:2024eag} we showed by direct Monte Carlo simulation that the SP regime is the most robust, and a simple Knudsen correction was able to account for heat transport across all regimes, and for different interaction types. Below, we briefly review this formalism, and the Knudsen correction. 

In the SP, non-local regime ($K\gg 1$), the DM population is modelled by an isothermal temperature profile with a constant temperature $T_\chi$. This temperature represents the average stellar temperature sampled by the DM population through its interactions with the star weighted by the radius-dependent collision rates. In this regime, the DM distribution is given by 

\begin{equation}
n_{\chi}(r) = N_{\text{iso}} e^{-m_\chi \: \phi (r) / T_\chi},
\end{equation}
where $m_\chi$ is the DM mass, $\phi(r)$ is the potential at radius $r$, and $N_{\text{iso}}$ normalises the distribution to the total number of DM particles in the star. 

Assuming the DM has uniform temperature $T_\chi$, scattering with target nuclei or electrons with number density $n_T(r)$ and temperature $T(r)$ at each radius $r$, yields an energy transport rate of:
\begin{equation}\label{equation:energyTransportFinalConst}
\epsilon(r) = \frac{8}{\rho(r)}\sqrt{\frac{2}{\pi}} \frac{m_\chi m_T}{(m_\chi+m_T)^2} n_\chi (r) n_T (r) 2\sigma_{0} \left( T_\chi - T(r) \right) \left( \frac{T(r)}{m_T} + \frac{T_\chi}{m_\chi} \right)^{\frac{1}{2}}.
\end{equation}
We present the full derivation of this equation in Appendix \ref{sec:heattransportderivation}, along with the equivalent expressions for momentum and velocity-dependent scattering. The luminosity at radius $r$ is then given by 
\begin{equation}\label{eq:lum_v_r}
L(r) = 4 \pi \int_0^{r} \rho (r^\prime) r^{\prime 2} \epsilon (r^\prime, T_\chi) dr^\prime.
\end{equation}
The isothermal temperature sets the sign for the energy transport. For the region where $T_\chi < T(r)$, $\epsilon$ is negative and so DM carries energy away, resulting in cooling of this region. This occurs in the inner layers of the star where the temperature is high. Likewise, when $T_\chi > T(r)$, the energy transport is positive and so DM deposits energy and heats up these regions.

The isothermal DM temperature $T_\chi$ is determined by demanding that there is no net energy loss or gain, leading to the condition
\begin{equation}\label{eq:lumTot}
L_{\text{tot}} = 4 \pi \int_0^{R_\star} \rho (r) r^2 \epsilon (r, T_\chi) dr = 0~.
\end{equation}
In practice, this is solved iteratively to find $T_\chi$.

As mentioned above, this must be corrected when moving away from the isothermal regime, when interactions are stronger. We use the simple prescription derived and verified in Refs. \cite{Banks:2021sba,Banks:2024eag}:
\begin{equation}\label{equation:modificationFactor}
L_{\text{calib}} = \frac{0.5}{1+\left( K_0/K \right)^2} L_{\text{SP}}, 
\end{equation}
where the value of $K_0$ depends mostly on the interaction type and the target type, and has a small dependence the DM mass. For $K \gg K_0$ values (SP regime), the modification factor reduces to a scaling of $0.5$ and so $L \propto \sigma_0$. While for $K \ll K_0$ values (LTE regime), the modification factor suppresses the luminosity and grows as $\sim K^2$, and so $L \propto \sigma_0^{-1}$.

In this work, we use the modified SP formalism to model the heat transport across $K$ values. We take an approximate value of $K_0 \simeq 0.4$ which was found by Refs. \cite{Banks:2021sba,Banks:2024eag} to be a good fit for constant DM-nucleon interactions.
\begin{figure}
    \centering
    \includegraphics[width=0.5\textwidth]{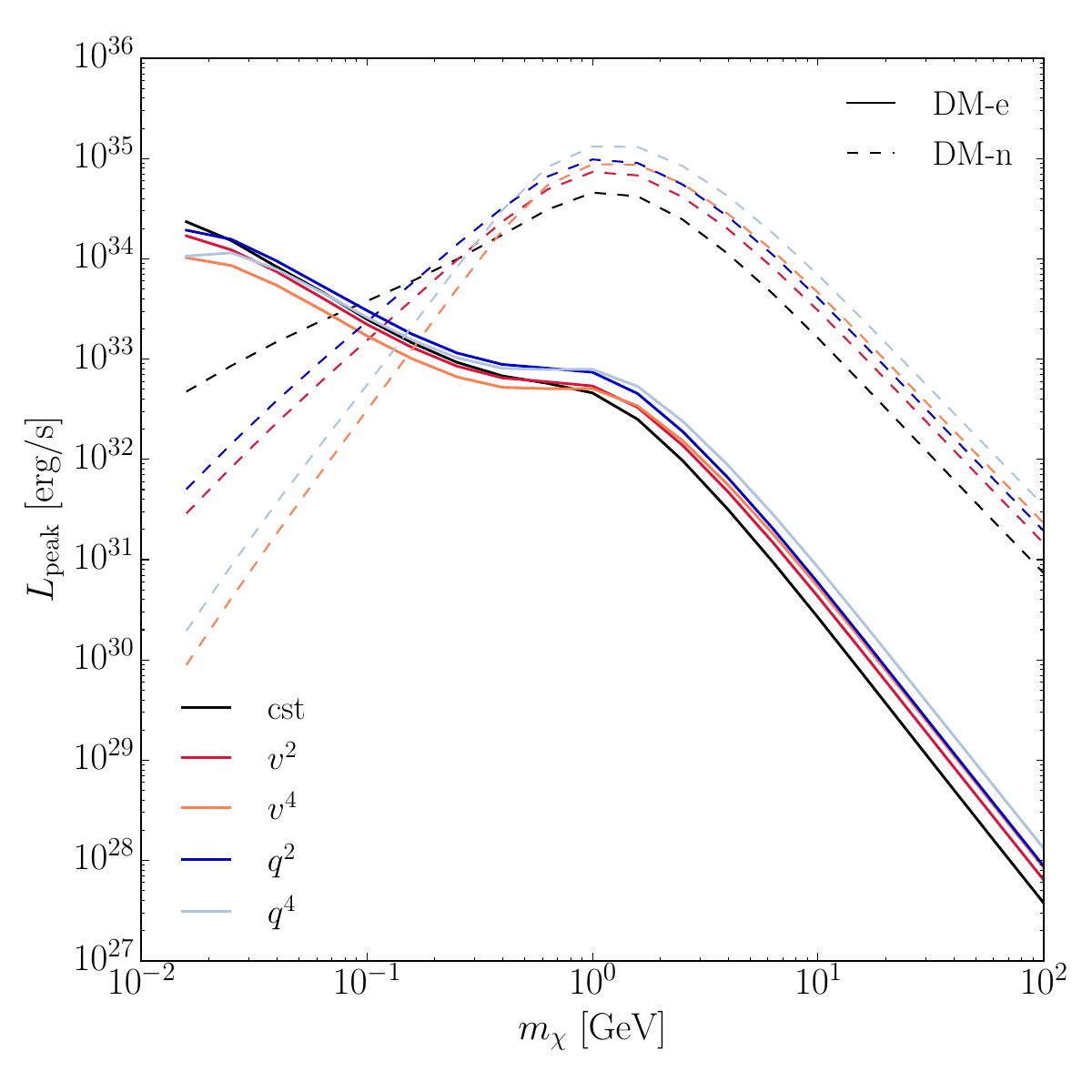}\includegraphics[width=0.5\textwidth]{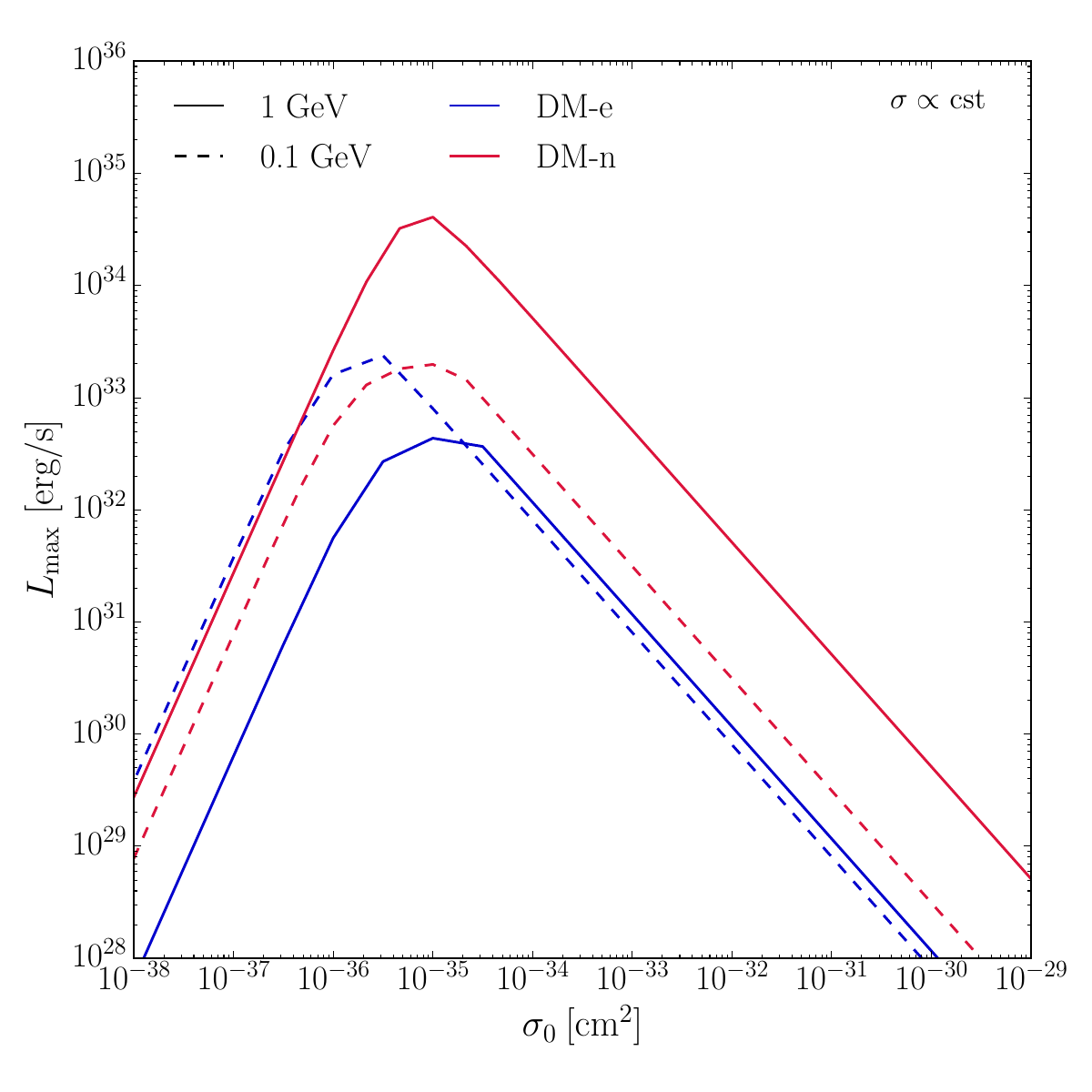}
    \caption{Peak transported luminosity by asymmetric dark matter in a standard solar model as a function of dark matter mass (left) and cross section (right), accounting for the computed capture rate.}
    \label{fig:lumvsigma}
\end{figure}
The resulting luminosity is shown as a function of mass and cross section in Fig.~\ref{fig:lumvsigma} for DM-nucleon  and DM-electron interactions in a present-day Standard Solar Model. For each cross section, the luminosity is computed with a DM population resulting from the expected capture rate \eqref{DM_Capture_Rate}, but without accounting for evaporation. The left panel shows the luminosity as a function of DM mass. The peak just above 1 GeV for DM-nucleon interactions is due to kinematic matching with hydrogen. The corresponding peak for DM-electron interactions would be near the electron mass, in which case the DM would not be confined to the star. It is tempting to attribute the shoulder near 1 GeV as being related to this peak; however, its presence is due to the extent of DM cloud at these temperatures, which brings most of the deposited heat right outside the stellar core. Its presence is therefore more reflective of the stellar structure and the equilibrium distribution of the DM, than of any microphysics. Different colours show the result for cross sections that scale as $v^n$ or $q^n$. The right panel of \ref{fig:lumvsigma} shows equivalent plots, but as a function of the DM-target cross section, clearly illustrating the Knudsen peak. Its location is slightly shifted, and the slope of the $L_{\rm max}$ curve is slightly steeper than would be suggested by Eq. \eqref{equation:modificationFactor} (and as compared to e.g. Fig. 3 of \cite{Banks:2021sba}) because these curves are again weighted by the capture rate. Figure \ref{fig:lumvnqn} in Appendix \ref{sec:appendix} shows the equivalent curves for non-constant cross sections.

Using the SP formalism offers the advantage of providing an analytic expression for the radial energy transport profile, simplifying computations. In contrast, the GR formalism requires calculating diffusion and conduction coefficients and numerically differentiating stellar models to determine the dark matter distribution and luminosity, making implementation more challenging and less stable. Indeed, the resulting numerical instabilities have led to incorrect or ambiguous results in the literature.

So far, we have laid out the theory and equations needed to calculate the capture rate of DM in a star during its evolution and the heat transport caused by the DM population. Next, we turn to stellar modelling and asteroseismology. 

\subsection{Effects on stellar structure and asteroseismology}
The pp chain, responsible for most heat produced in stars below $\sim 1.2 M_\odot$, scales with the temperature as $T^4$. In heavier stars, the CNO cycle dominates, scaling as $\sim T^{20}$.
Such stars develop a steep temperature gradient, leading to a breakdown of local hydrostatic equilibrium and the development of convective motion in the core. These convective cores lead to uniform mixing of hydrogen into the burning region, and a sharp boundary that can be inferred from asteroseismology. 

The oscillation frequencies of a star are determined from time series observations of their line-of-sight velocity. Once the oscillation frequencies are observed, asteroseismology aims to extract information about the star's interior using an inversion method. This method involves adjusting the structure of a reference model to match the observed frequencies. The main idea is to treat the stellar oscillations as linear perturbations around a known stellar model, adjust model parameters to minimize the difference between the modelled and observed quantities, and consequently obtain information about the stellar structure. 
Through the use of different inversion methods, we can deduce different structural information about the star such as the sound speed profiles, surface Helium abundance, core composition. 

Individual model oscillation frequencies are sensitive to surface effects and the modeling of a star's surface layers, which are poorly described even for solar models. However, certain ratios of frequency combinations are independent of the surface layer details and are therefore more reliable parameters to describe the stellar interior \cite{roxburgh2003ratio}.

One commonly used set of ratios for stellar modeling, due to its sensitivity to the stellar core properties is \cite{roxburgh2003ratio,Basu:2006vh,Chaplin:2007uh}: 
\begin{equation}\label{eq:r02}
    r_{02} (n) = \frac{d_{02}(n)}{\Delta_1(n)},
\end{equation}
where $d_{02}(n)$ and $\Delta_1(n)$ are the small and large frequency separation ratios given by
\begin{equation}
    d_{l, l+2}(n) = \nu_{n, l} - \nu_{n-1, l+2} \simeq -\left( 4l + 6 \right) \frac{\Delta_l(n)}{4\pi^2 \nu_{n, l}} \int_0^{R_\star} \frac{dc_s}{dr} \frac{dr}{r} ,
\end{equation}
and
\begin{equation}\label{eq:largefreqsep}
    \Delta_{l}(n) = \nu_{n+1, l} - \nu_{n, l} \simeq \left( 2 \int_0^{R_\star} \frac{dr}{c(r)} \right)^{-1}, 
\end{equation}
respectively, where $R_\star$ is the star's radius, $c_s(r)$ is the sound speed profile, and $\nu_{n,l}$ is the oscillation frequency. 

\section{Dark Matter in Main-Sequence Stars}\label{section:other_star}

In this section, we explore the impact of DM heat transport on the structure and evolution of stars and illustrate the changes it induces for both interactions with electrons and nucleons. In Sec. \ref{sec:realstar} we will focus on applying this analysis to a nearby star for a range of possible dark matter masses and cross sections. 

For stellar modelling we use release 15140 of \texttt{MESA} (Modules for Experiments in Stellar Astrophysics), an open-source 1D stellar evolution code \cite{2011ApJS..192....3P,2013ApJS..208....4P,2015ApJS..220...15P,2018ApJS..234...34P,2019ApJS..243...10P,2023ApJS..265...15J}. \texttt{MESA} allows users to model a star given a set of initial stellar parameters. Capture and heat transport are modeled by \texttt{Capt'n General} \cite{kozar2021capt}, which we have   adapted to include DM-electron scattering, and to calculate the capture rates at finite target temperature as discussed in Sec. \ref{DM_Capture_Rate}.  To include the impact DM has on stars we use the \texttt{MESA} provided \texttt{extra\_energy\_implicit} hook. We interface with \texttt{Capt'n Gen} and provide it with the stellar profiles and properties and in return \texttt{Capt'n Gen} calculates the energy removed or deposited by dark matter as a function of stellar radius. \texttt{MESA} then includes this extra energy contribution into its stellar structure equations and evolves the model accordingly. 

We will consider stars of three distinct masses: one that would normally lead to a radiative core, one resulting in a convective core, and another at the transition between these two.

Our focus in this section is on observing general trends resulting from the inclusion of DM in stellar evolution, thus we do not perform any precise stellar calibrations. We choose fiducial values for the initial parameters of the star in all our stellar modelling with and without DM. We use an initial metallicity of $[\text{Fe/H}]=-0.1$, initial helium abundance $Y_i = 0.22$, and mixing length parameter $\alpha = 1.5$. We evolve our stars until their core hydrogen abundance is half of its initial value $X = 0.5 \ X_{i}$. 
We use local values for the stellar velocity $v_\star = 220 $ km~s$^{-1}$ , DM velocity dispersion $v_0 = 270$ km~s$^{-1}$  and DM density $\rho_{\chi} = 0.4$ GeV~cm$^{-3}$.

\begin{figure}
    \centering
        \includegraphics[width=0.8\linewidth]{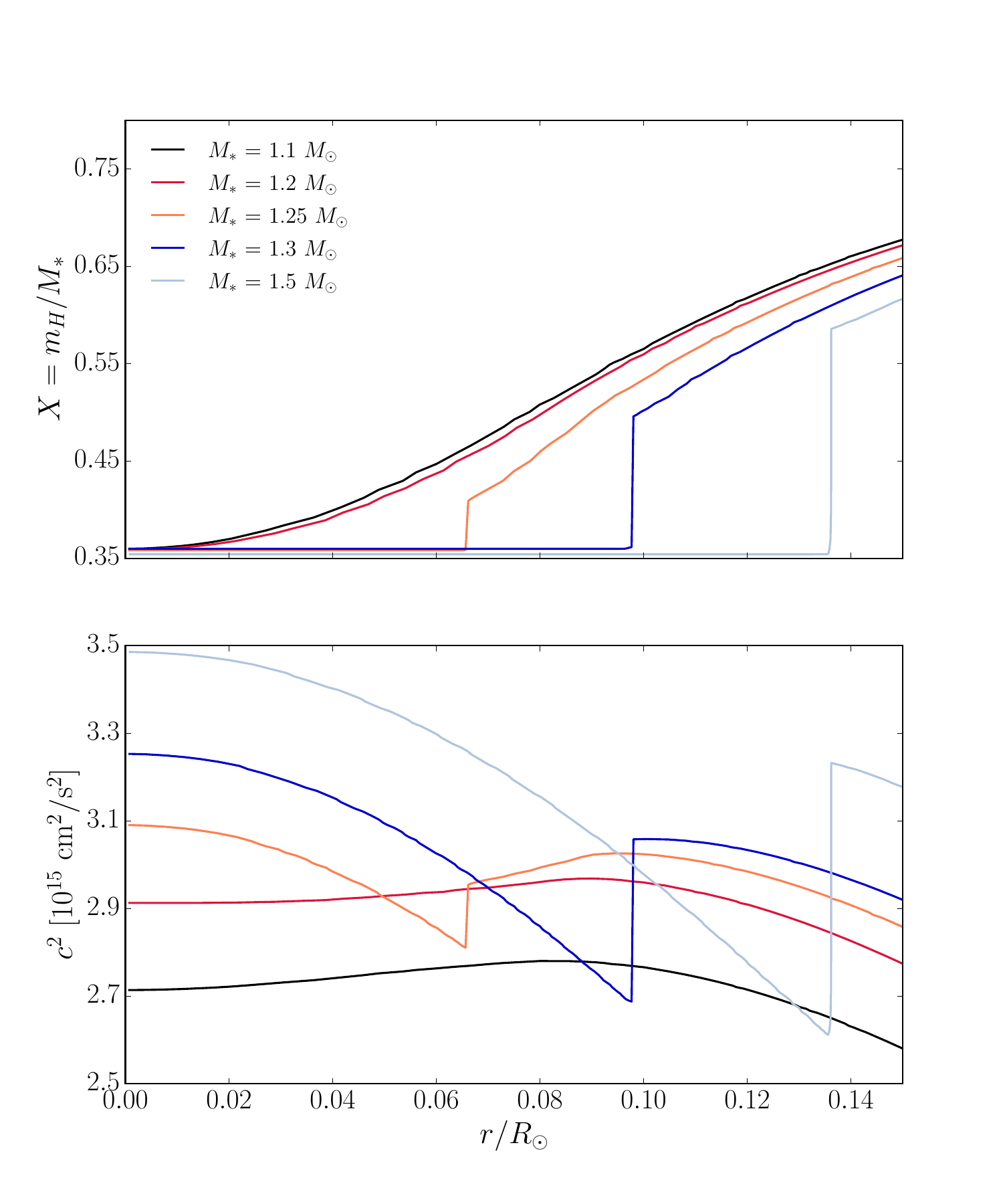}
     \caption[Hydrogen mass fraction and squared sound speed profile for stars with varying mass evolved until their central hydrogen mass fraction is half its initial value without the presence of DM.]{\textit{Hydrogen mass fraction (upper panel) and squared sound speed profile (lower panel)} for stars with varying masses evolved until their central hydrogen mass fraction is half its initial value without the presence of DM.}
    \label{fig:noDM_conv_core_develop}
\end{figure}

We begin by determining the mass at which stars transition from having a radiative to a convective core under our specified initial conditions.
We run stellar simulations without DM for stars with masses between $1.1 \ M_\odot$ and $1.5 \ M_\odot$ with a mass step of $0.05 \ M_\odot$ to determine at which mass the convective core develops.
Figure \ref{fig:noDM_conv_core_develop} shows the radial profiles of the hydrogen mass fraction $X$ (top panel) and the square of the sound speed $c^2$ (bottom panel) for stars with masses of 1.1, 1.2, 1.25, 1.3, and 1.5 $M_\odot$. 
Based on changes in these profiles, we identify that the mass threshold for developing a convective core is $1.25 \ M_\odot$ for our initial parameters.
Below this mass, $X$ increases smoothly from the core outward, while above it, $X$ remains constant up to a certain radius before sharply increasing, indicating the presence of a convective core. The lower panel of Figure \ref{fig:noDM_conv_core_develop} shows the resulting discontinuity  in the sound speed $c(r)$ due to the resulting abrupt change in mean molecular weight. 

Next, we integrate DM capture and energy transport into the stellar evolution. We evolve three stars with masses of 1.1, 1.25, and 1.5 $M_\odot$ until their core hydrogen mass abundance $X$ is half its initial value $X_i$ while accounting for the impact of DM. 

\begin{figure}[h]
        \begin{center}
            \includegraphics[width=1.0\linewidth]{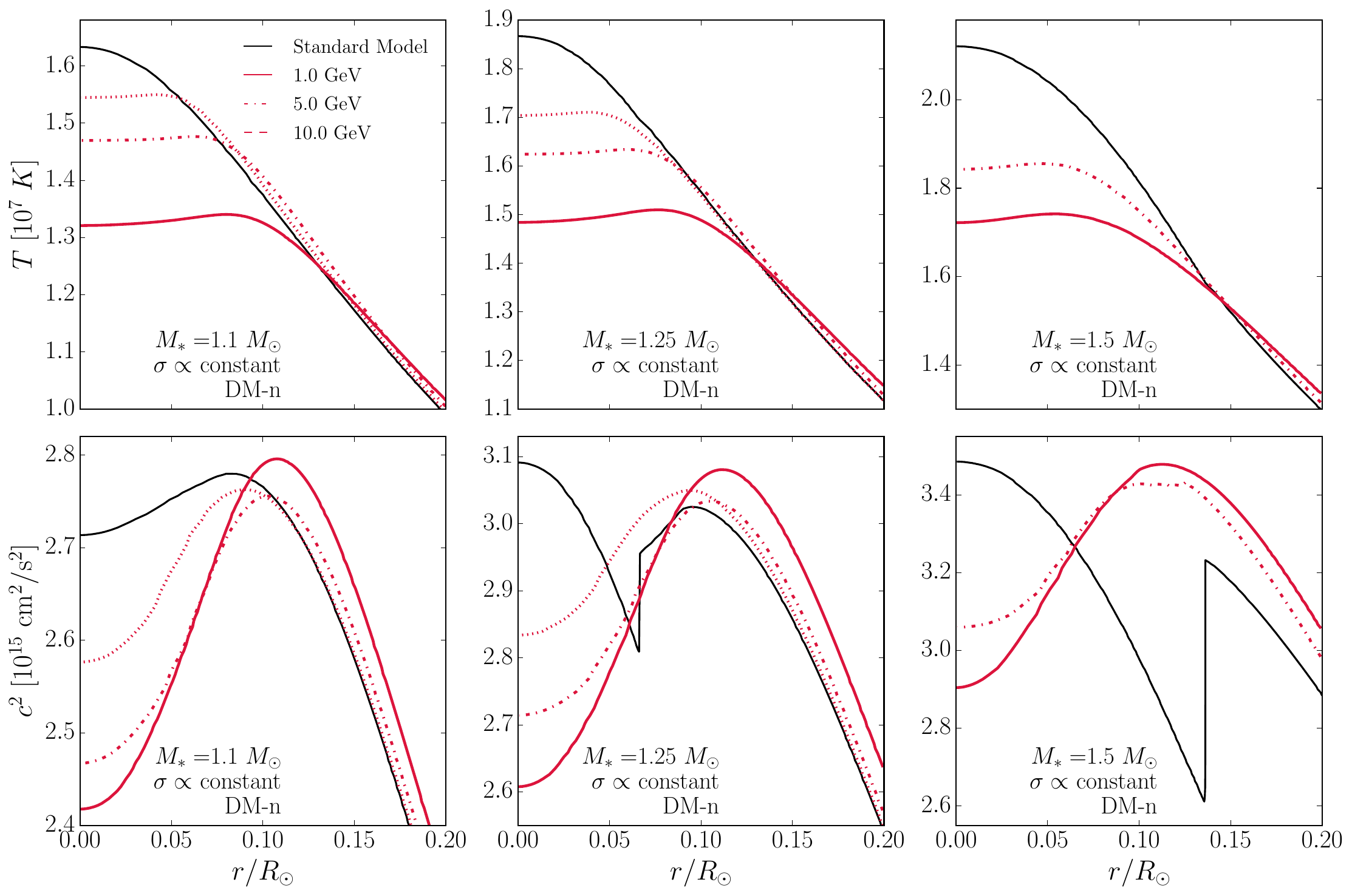}
        \end{center}
         \caption[The temperature  and squared sound speed profiles for $1.1 \ M_\odot$, $1.25 \ M_\odot$ , and $1.5 \ M_\odot$ stars with constant DM-nucleon interactions.]{\textit{The temperature (upper panels) and squared sound speed (lower panels) profiles} for $1.1 \ M_\odot$ (left), $1.25 \ M_\odot$ (middle), and $1.5 \ M_\odot$ (left) stars without DM (black curves) and with constant DM-nucleon interactions (red curves) for 1 GeV (solid lines), 5 GeV (dotted lines), and 10 GeV (dashed-dotted lines) DM.}
            \label{fig:radialProfiles_DM-hydrogen}
\end{figure}

Radial temperature profiles resulting from DM-nucleon interactions are shown in Figure \ref{fig:radialProfiles_DM-hydrogen} for a 1.1 $M_\odot$ (left), 1.25 $M_\odot$  (middle) and 1.5 $M_\odot$  (right) star. Solid, dash-dotted and dotted lines respectively show the effects of a 1, 5, and 10 GeV DM particle. The cross section in each case is chosen to be at the Knudsen peak, i.e. where heat transport is maximised. The 10 GeV case for DM-nucleon and $1.5 \ M_\odot$ is omitted as it did not converge under our initial parameter choices and did not reach the stopping condition.

For $M = 1.1 \ M_\odot$, DM of mass 1, 5, and 10 GeV reduce the central temperature of the star by 20 \%,  10 \%, and 5 \%, respectively. For the $1.25 \ M_\odot$ star, DM of mass 1, 5, and 10 GeV reduce the central temperature of the star by 20 \%,  13 \%, and 9 \%, respectively, with all three masses erasing the convective core. 
For the more massive star with a convective core, $M = 1.5 \ M_\odot$, DM of mass 1 GeV and 5 GeV were also able to erase the convective core.

\begin{figure}[h]
        \begin{center}
            \includegraphics[width=1.0\linewidth]{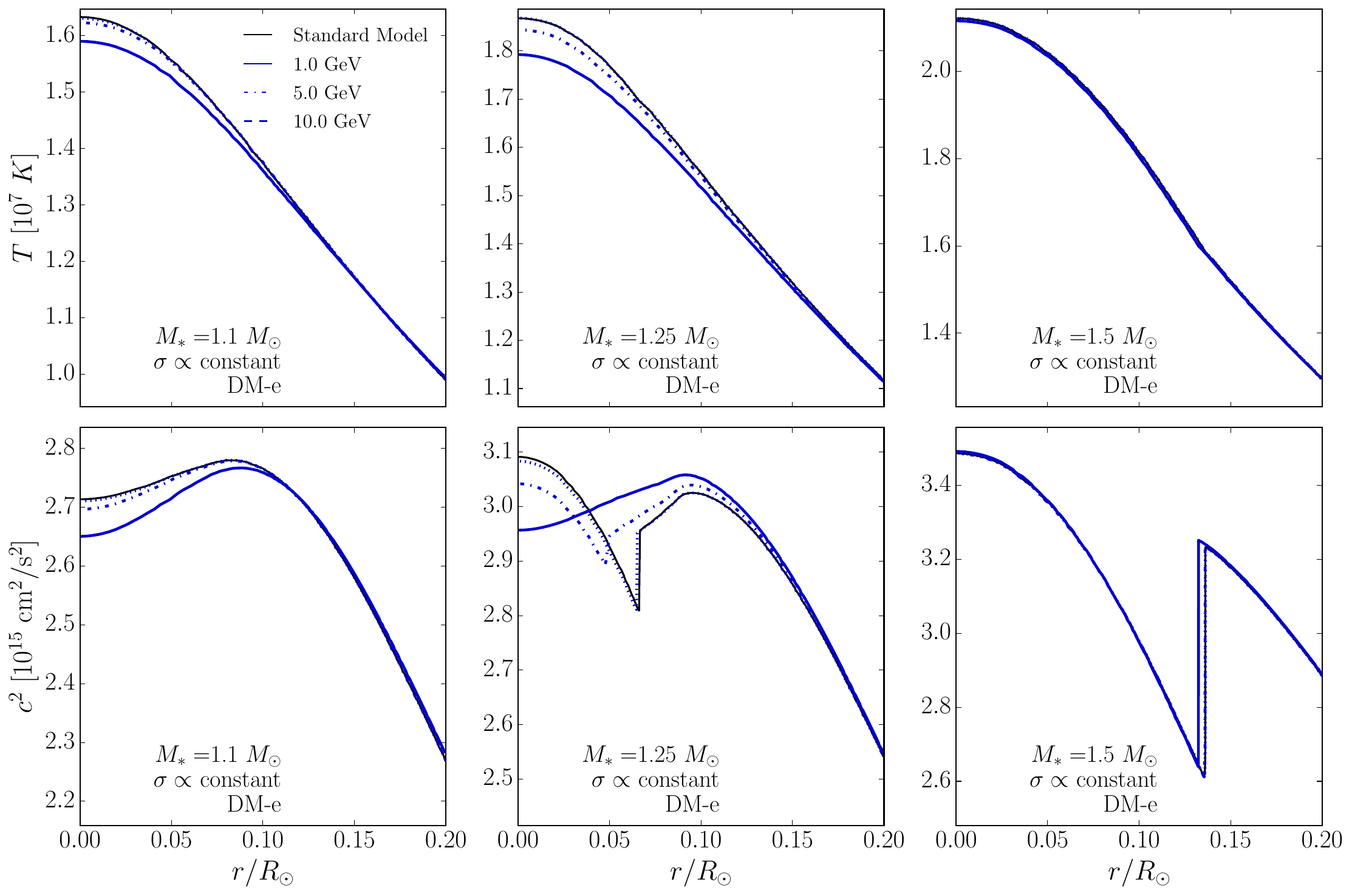}
        \end{center}
         \caption[The temperature  and squared sound speed profiles for $1.1 \ M_\odot$, $1.25 \ M_\odot$ , and $1.5 \ M_\odot$ stars with constant DM-electron interactions.]{\textit{The temperature (upper panels) and squared sound speed (lower panels) profiles} for $1.1 \ M_\odot$ (left), $1.25 \ M_\odot$ (middle), and $1.5 \ M_\odot$ (left) stars without DM (black curves) and with constant DM-electron interactions (blue curves) for 1 GeV (solid lines), 5 GeV (dotted lines), and 10 GeV (dashed-dotted lines) DM.}
            \label{fig:radialProfiles_DM-Electron}
\end{figure}

 In Figure \ref{fig:radialProfiles_DM-Electron} shows the equivalent set of plots, but for DM-electron interactions. For stars with a radiative core, $M = 1.1 \ M_\odot$, DM masses of 1 GeV and 5 GeV reduce the central temperature by 3 \% and 1 \%, respectively, while DM of mass 10 GeV shows no significant effect compared to the no-DM case.
For the $1.25 \ M_\odot$ star, DM of mass 1 GeV lowers the core temperature  by 4 \% and erases the convective core as evidenced in the changes in the radial $c^2$ profile compared to the no-DM case.
As for $m_\chi = 5 \ \text{GeV}$, DM reduces the core temperature by 1 \%, but fails to erase the convective core of the star, although it does shrink it in size. For $m_\chi = 10 \ \text{GeV}$, no significant impact is observed inside the star, aligning with the no-DM scenario. For more massive stars with a convective core, $M = 1.5 \ M_\odot$, constant DM-electron interactions do not notably alter the star's properties.

It is evident that DM-nucleon interactions have a larger impact on the stars. This is due to the consistently higher capture rate and peak of maximum luminosity for $m_\chi \gtrsim 1 \ \text{GeV}$ (Figs \ref{fig:capratev0q0} and \ref{fig:lumvsigma}), thanks in turn to the better kinematic matching with nucleons. For both the DM-nucleon and DM-electron interactions, the impact of DM on the star diminishes with increasing DM mass as expected.

The impact of DM on stars with a convective core decreases as stellar mass increases. A more massive star has a larger convective core and a steeper temperature gradient, requiring a more substantial impact from DM to affect their radial profiles, even within the core. We could increase the impact of DM by considering regions with larger DM densities ($\rho_\chi > \rho_\odot$) which would increase the total luminosity without increasing the suppression factor (given by Equation \ref{equation:modificationFactor}) which is used when modelling the energy transport. However because our ultimate target is a nearby star, we limit ourselves to local regions where $\rho = 0.4$ GeV cm$^{-3}$. 

It is evident that stars with masses around $1.25 \ M_\odot$ represent compelling targets, particularly for studying DM-electron interactions. For instance, while both $1.1 \ M_\odot$ and $1.25 \ M_\odot$ stars experience a similar decrease in central temperature due to DM-electron interactions, the impact on stellar profiles is much more pronounced for stars just above the threshold to develop convective cores. 

The erasure of the convective core also impacts the evolution of the star. In scenarios without DM, the convective core extends beyond the hydrogen-burning region and therefore hydrogen from the outside of this burning region is mixed into it through convection. For ADM models which erase the convective core, the stellar core no longer gets an influx of hydrogen from outside the hydrogen burning zone and the star has less fuel available to burn, therefore the central hydrogen fraction decreases faster. This behaviour was illustrated in Ref. \cite{raen2021effects}.

\begin{figure} 
        \centering
        \includegraphics[width=0.5\textwidth]{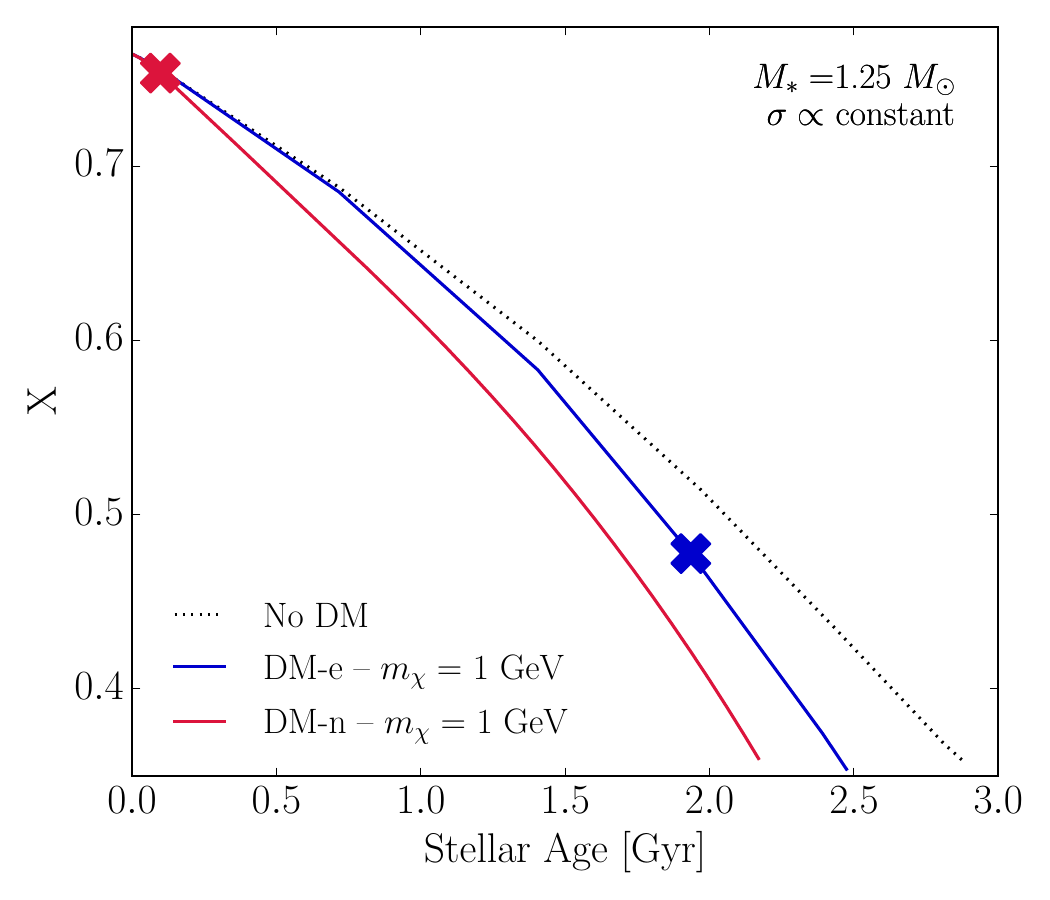}\includegraphics[width=0.5\textwidth]{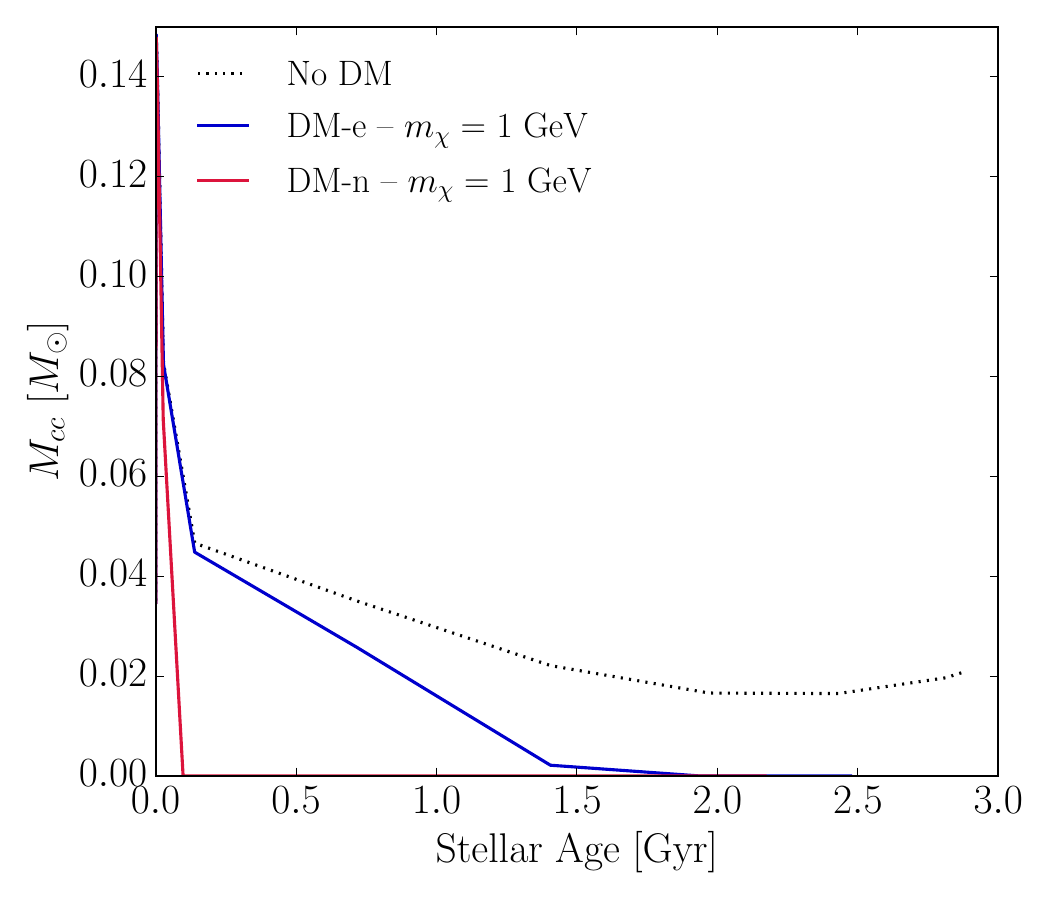}
     \caption[The evolution of the central hydrogen mass fraction and the convective core mass for a $1.25 \ M_\odot$ star with constant DM-electron and DM-nucleon interactions for DM mass of 1 GeV.]{\textit{Central hydrogen mass fraction (left panel) and the mass of the convective core (right panel) as a function of the stellar age} without DM (dotted black line), and with 1 GeV constant DM-electron (blue) and DM-nucleon (red) interaction. The blue and red cross marks show the age at which the convective core mass is zero for the DM-electron and DM-nucleon case, respectively.}
     \label{fig:X_and_Mc_v_age_electrons_hydrogen}
\end{figure}

Figure \ref{fig:X_and_Mc_v_age_electrons_hydrogen} illustrates the central hydrogen abundance (left panel) and the mass of the convective core (right panel) over the stellar age without DM (dashed black line) and with DM-electron (blue) and DM-nucleon (red) constant interactions with a DM mass of 1 GeV. The blue and red cross marks show the point at which the convective core is completely erased for interactions with electrons and nucleons, respectively.

DM-nucleon collisions erase the convective core at 0.1 Gyr. In the absence of a convective core, the hydrogen in the core is depleted more rapidly relative to the no-DM case.
For the DM-electron case, the convective core is erased at 2 Gyr. Even before its erasure, the presence of DM accelerates the reduction rate of the central hydrogen fraction $X$ relative to the no-DM case after 0.8 Gyr. This acceleration is due to DM gradually reducing the size of the convective core before erasing it, as shown in the right panel of Figure \ref{fig:X_and_Mc_v_age_electrons_hydrogen}. Because DM-nucleon interactions are more efficient than DM-electron interactions, they erase the convective core at a much earlier age than DM-electron interactions, thus leading to a larger impact on the stellar evolution. 

The impact of DM on the star varies depending on the type of interaction (constant and $v/q$ dependent), but the general trends are qualitatively similar. 

\begin{figure}
        \begin{center}
            \includegraphics[width=0.6\linewidth]{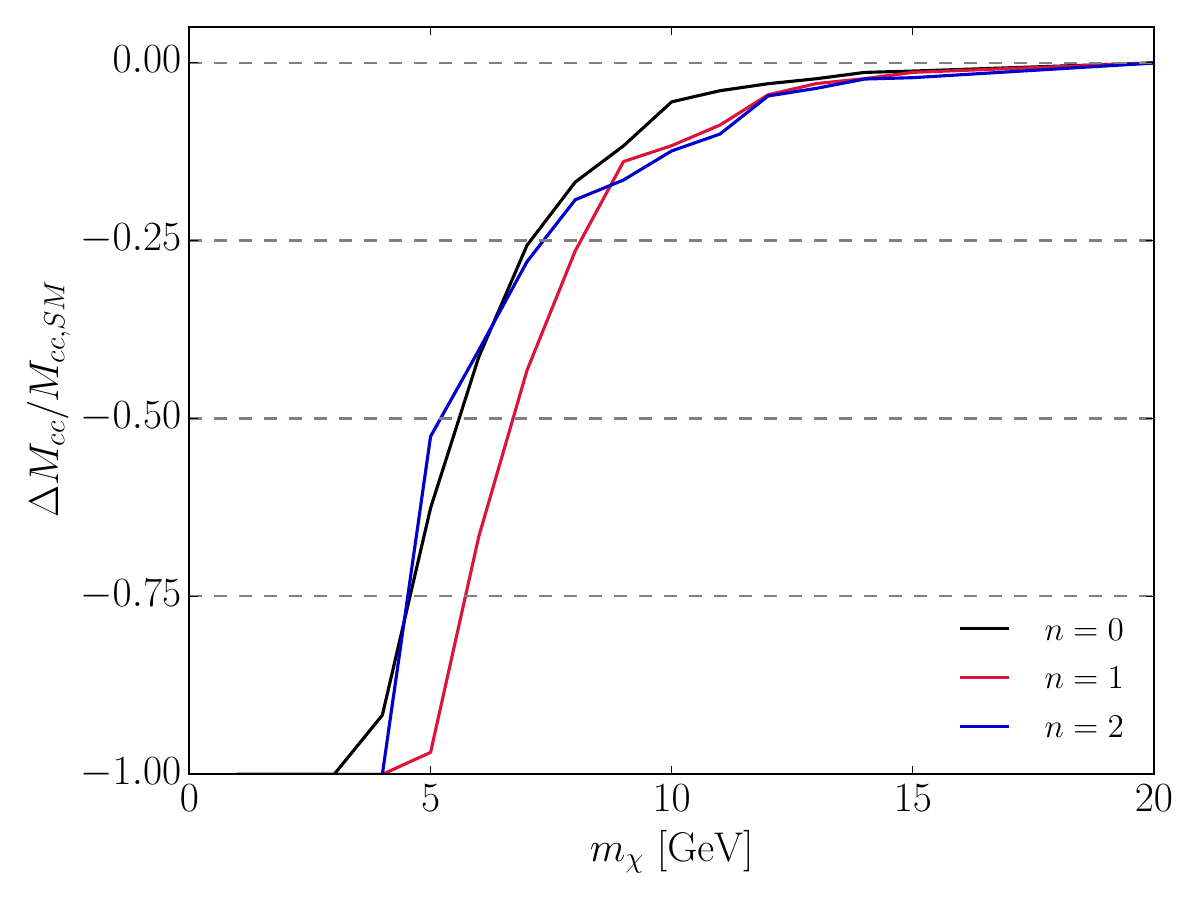}
            \caption[Change in the mass of the convective core of a $1.25 \ M_\odot$ star due to the presence of DM for DM-electron interactions.]{\textit{The change in the mass of the convective core of a $1.25 \ M_\odot$ star due to the presence of DM relative to the no-DM case as function of the DM mass} for DM-electron constant (black), $n=1$ (red), $n=2$ (blue) interactions where the cross section is chosen such that the effect is maximized at each given mass. The star is evolved until $X = 0.5X_i$.}
            \label{fig:delta_Mcc_electron}
        \end{center}
\end{figure}

In Figure \ref{fig:delta_Mcc_electron}, we show the maximum (that is, evaluated at the value of $\sigma_0$ where capture and transport are maximal) impact DM has on the convective core mass $M_{cc}$ of a $1.25 \ M_\odot$ star compared with the no-DM case for $n = {0, 1, 2}$ to denote the constant, $v^2/q^2$, and $v^4/q^4$ DM-electron interactions. This is represented as the relative change in the mass of the convective core $\Delta M_{cc}$ compared to the no-DM case. 
The identical behaviours for $v^{2 n_v}$ and $q^{2 n_q}$ with $n_v = n_q = n$ are due to the fact that their heat transport profiles only differ by a constant, which can be removed by scaling $\sigma_0$. 

There is a complete erasure in the convective core up to $m_\chi = $ 3, 4, and 4 GeV for the $n=0$, $n=1$, and $n=2$ scenarios, respectively. Beyond these masses, DM shrinks the size of the convective core up until $m_\chi \approx 10 \ \text{GeV}$.
For DM masses $> 10 \ \text{GeV}$, DM tracks the no-DM case with the resulting convective core having the same mass as the no-DM case. The $n=0, 1, 2$ DM-electron interactions show comparable behaviours, with $n=1$ interactions having a slightly more significant impact on the convective core compared to the $n=0$ and $n=2$ interactions.

\begin{figure} 
    \begin{subfigure}
        \centering
        \includegraphics[width=0.5\linewidth]{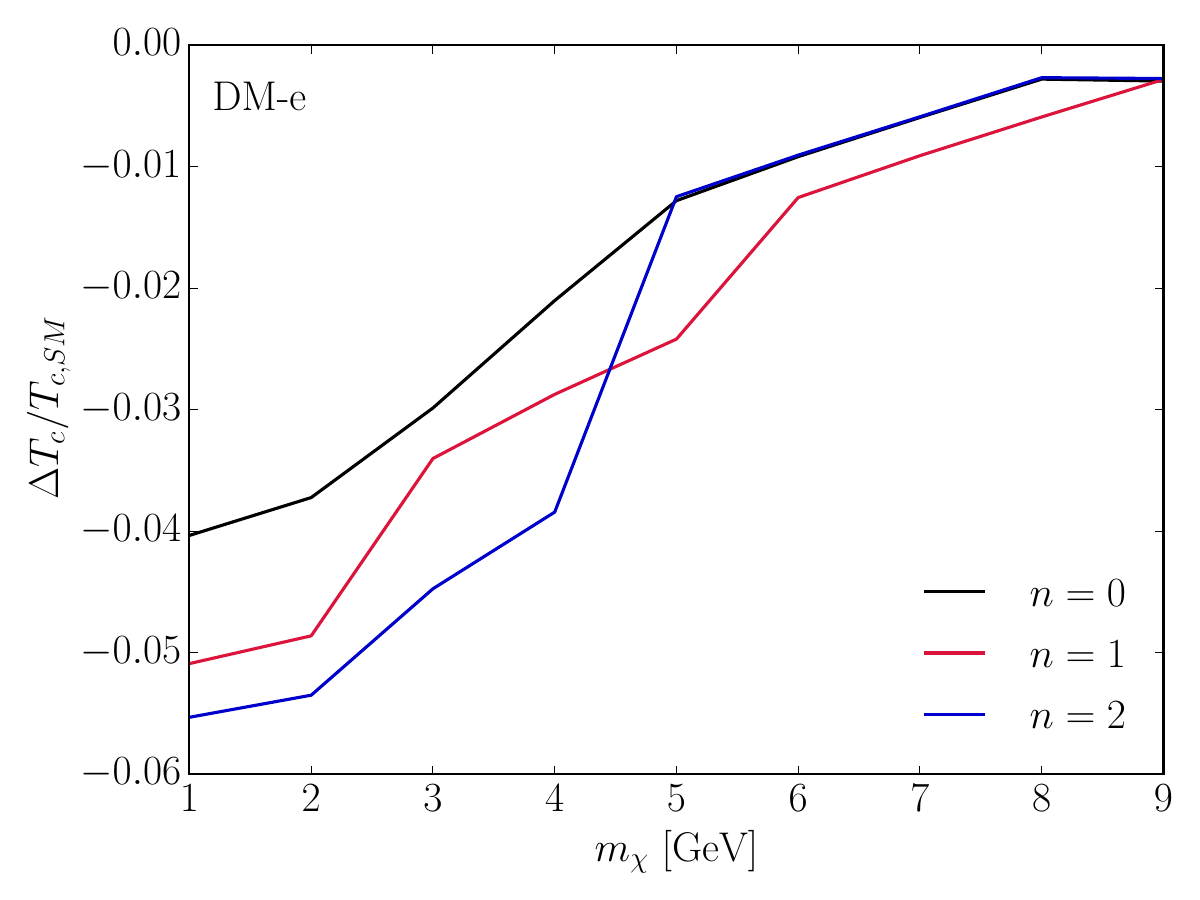}
    \end{subfigure}%
    \begin{subfigure}
        \centering
        \includegraphics[width=0.5\linewidth]{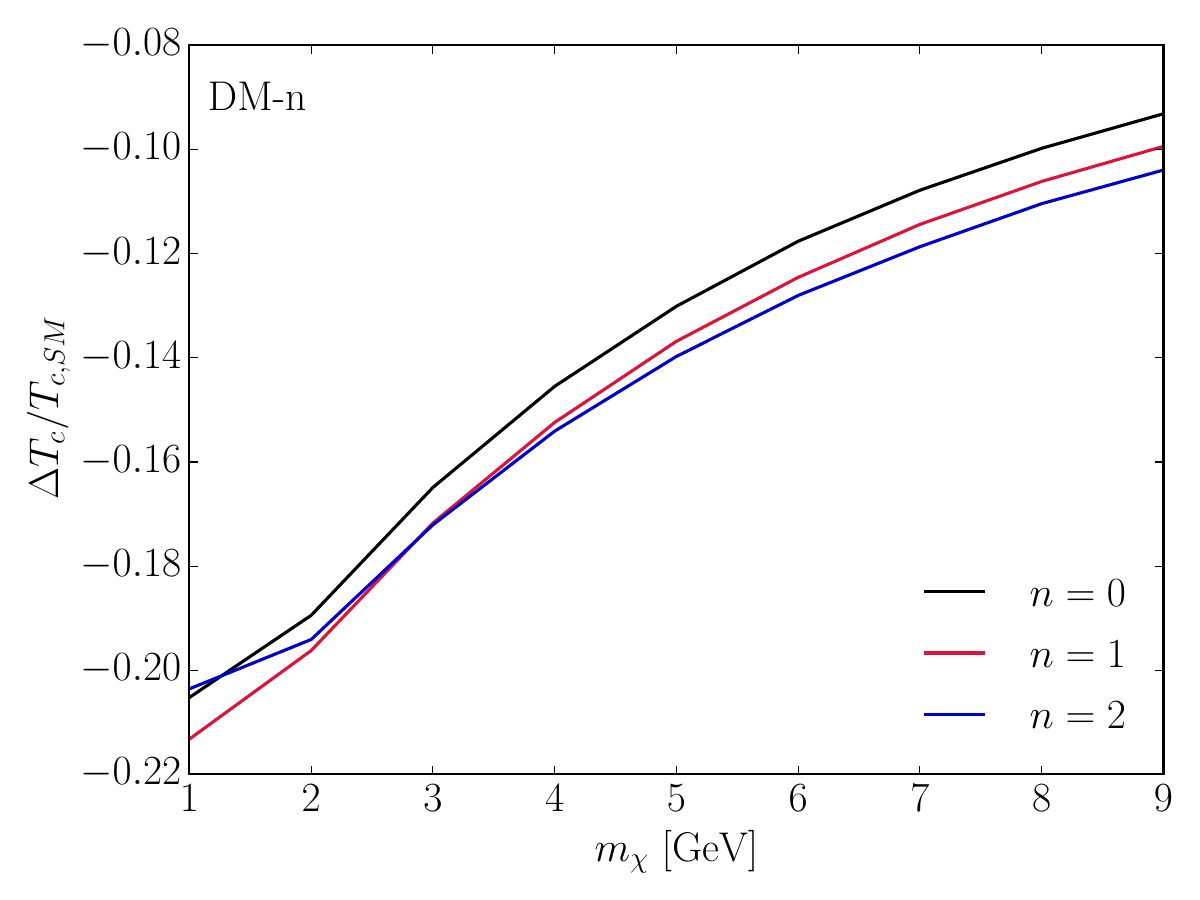}
    \end{subfigure}
     \caption[The change in the core temperature of a $1.25 \ M_\odot$ star due to the presence of DM for DM-electron interactions.]{\textit{The change in the core temperature  relative to the no-DM case of a $1.25 \ M_\odot$ star as function of the DM mass} for DM-electron (left panel) and DM-nucleon (right panel) constant (black), $v^2/q^2$ (red), $v^4/q^4$ (blue) interactions where the cross section is chosen such that the effect is maximized at each given mass. The star is evolved until $X = 0.5X_i$.}
     \label{fig:Delta_Tc_electrons_hydrogen}
\end{figure}

In Figure \ref{fig:Delta_Tc_electrons_hydrogen}, we show the reduction in the core temperature $\Delta T_c$ relative to the no-DM case for DM-electron (left) and DM-nucleon (right) interactions. As expected, the greatest reduction in core temperature occurs for 1 GeV DM for both DM-electron and DM-nucleon interactions, and decreases as the DM mass increases.
Specifically, for DM-electron interactions, the core temperature reduction is 4, 5, and 5.5 \% for $n=0$, $n=1$, and $n=2$ interactions, respectively for 1 GeV DM. As the mass reaches 9 GeV, its impact on core temperature becomes negligible.
In contrast, DM-nucleon interactions with 1 GeV DM decrease the core temperature by approximately 20 \% for $n=0, 1, 2$ interactions. Even at 9 GeV, the reduction (about 10 \%) remains larger than that observed for 1 GeV DM-electron interactions.

\section{Dark matter in KIC 8228742}
\label{sec:realstar}
In the previous section, we investigated the impact of ADM on stars of different masses. We highlighted the importance of convective cores and demonstrated how DM-electron and DM-nucleon interactions can eliminate these cores, leading to significant alterations in the stellar radial profiles and evolution. Here, we turn to effects on a specific star, and compare predictions to observation.

\subsection{Stellar calibration}
\label{sec:calibration}
We will use the \texttt{MESA astero} module for stellar calibration, which interfaces with both GYRE \cite{townsend2013gyre} and ADIPLS \cite{christensen2008adipls} for asteroseismic calculations, enabling the computation of pulsation frequencies throughout the stellar model's evolution.
Stellar calibration involves adjusting input parameters iteratively in a simulated stellar model to match observational constraints closely. This iterative process continues until we achieve a model that fits the observational data best. Stellar calibration is crucial when adding new physics,  ensuring that other physical parameters are self-consistently taken into account. 

The observational constraints used in the calibration may include spectroscopic observables (e.g. effective temperature $T_\text{eff}$, and the luminosity $L$) as well as asteroseismic constraints such as large frequency separation $\Delta \nu$ (Eq. \ref{eq:largefreqsep}) and individual frequencies. The initial parameters for the stellar modelling are the stellar mass $M$, the initial helium abundance $Y_i$, the initial metallicity $[\text{Fe/H}]_i$, the mixing length parameter $\alpha_{\text{MLT}}$, and the convective overshoot parameter $f_{\text{ov}}$. 

The \texttt{astero} module performs the calibration by minimizing a chi-squared $\chi^2$ statistic: 
\begin{equation}
    \chi^2 = \frac{1}{N} \sum_{i=1}^N \left( \frac{X_i^{\text{mod}} - X_i^{obs}}{\sigma_i} \right)^2, 
\end{equation}
where $N$ is a normalization factor given by the number of parameters, $X_i^{\text{mod}}$ is the simulated model's value for the $i$-th observable, $X_i^{\text{obs}}$ is the observed value, and $\sigma_i$ is the uncertainty.  
There are several calibration methods implemented in the \texttt{astero} module. In this work, we use the downhill simplex algorithm \cite{ja1965simplex}. 

 We perform an initial calibration using the \texttt{astero} module to match three spectroscopic observables: ${L}$, $T_{\text{eff}}$ and the  surface gravity $\log g$ as well as one asteroseimic observable, the large frequency separation $\Delta \nu$.  The latter can be rapidly calculated by inverting the sound speed profile, rather than from the individual frequencies.

From this calibration, we obtain an initial $\chi^2_{\text{star}}$ given by 
\begin{equation}\label{equation:chi2Star}
    \chi^2_{\text{star}} = 1/3 \ \chi^2_{\text{spectro}} + 2/3 \ \chi^2_{\text{seismo}}~.
\end{equation}
We weigh $\chi^2_{\text{spectro}}$ by one-third and $\chi^2_{\text{seismo}}$ by two-thirds, typical values used in the literature \cite{2013ApJS..208....4P, Rato:2021tfc, metcalfe2012asteroseismology}. 

Once we have a minimized $\chi^2_{\text{star}}$, we run the calibrated model with the same initial parameters and obtain a $\chi^2_{r_{02}}$ for the frequency separation ratios $r_{02}$ (Equation \ref{eq:r02}) given by  

\begin{equation}
    \chi^2_{r_{02}} = \sum_{15}^{22} = \left[  \frac{r_{02}^{\text{obs}}(n) - r_{02}^{\text{mod}}(n)}{\sigma_{r_{02}}^{\text{obs}}(n)} \right]^2,
\end{equation}
where $r_{02}^{\text{obs}}(n)$ and $r_{02}^{\text{mod}}(n)$ are the frequency separation ratios computed using the observational and model frequencies, and $\sigma_{r_{02}}^{\text{obs}}(n)$ is the observed uncertainty computed using error propagation from the individual frequencies.

The decision to not include these frequency separation ratios in the first calibration is due to computational constraints. The $r_{02}$ calculation involves calculating the individual frequencies and doing so at each time step in the calibration is very time consuming.
Only $\Delta \nu$, which is calculated using an asymptotic approximation, is included in the initial calibration to account for asteroseismology when calibrating. 

\subsubsection{KIC 8228742}

Next, we explore the predicted impact of DM on a real star
and compare the results with observational data. Specifically, we look at KIC 8228742, currently in the subgiant (SG) stage of evolution, following the main-sequence phase, where hydrogen burning in the core has ceased. KIC 8228742 is classified as an F9IV-V spectral type star with a previously modeled mass of $\sim 1.27 \ M_\odot$ \cite{metcalfe2014properties}. 

We choose KIC 8228742 for two primary reasons: 1) Its modeled mass aligns with stellar masses predicted to exhibit a convective core in the absence of DM, yet susceptible to core erasure due to DM-electron and DM-nucleon interactions, as shown in the Section \ref{section:other_star}; and 2) This star has been used by \cite{Rato:2021tfc} to investigate the impact of constant SD DM-nucleon interactions, similar to our current approach. However, they did not conduct a full calibration across a grid of DM masses and cross-sections but rather relied on initial values from the no-DM calibrated model. As mentioned in Sec. \ref{sec:calibration}, it is important to re-calibrate for each set of DM model parameters considered, in order to properly account for the effects of nuisance parameters. In effect, this is equivalent to profiling over the set of standard stellar parameters.

We perform an initial calibration to obtain a $\chi^2_{\text{star}}$, given by Equation \ref{equation:chi2Star}. The spectroscopic and asteroseimologal data for KIC 8228742 are taken from \cite{chaplin2013asteroseismic} and \cite{appourchaux2012oscillation}, respectively. 
We will scan over a range of DM masses and cross sections and calibrate each point on that grid using the simplex optimization method provided by the \texttt{astero} module. These calibrations will yield a set of initial conditions for each $m_\chi-\sigma_0$ combination which minimizes the $\chi^2_{\text{star}}$. 

In this section, we will only examine constant scattering cross sections. As shown in the previous section, non-constant interactions give qualitatively similar results. Further, our calibration procedure is numerically expensive, limiting computational feasibility. 

For DM-nucleon interactions, we create a grid of cross sections informed by the analysis in Sec. \ref{section:other_star}, between $\sigma_0 = 10^{-39}$ and $10^{-35}$ cm$^2$ in steps of 0.5 dex, and DM masses ranging from 1 to 7 GeV in increments  of 1 GeV. Initially, larger values of $\sigma_0$, above the Knudsen peak, were also considered, but these points did not converge during the initial calibration process. 
For the DM-electron analysis, we scan across the same mass range, but  over cross sections in the range $\sigma_0 = 10^{-37}$ to $10^{-33}$ cm$^2$. 

Since KIC 8228742 is only 0.17 kpc away, we assume a DM density similar to that in the solar neighbourhood $\rho_{\chi} = 0.4 \ \text{GeV/cm}^3$. We choose a stellar velocity of 220 km/s and a velocity dispersion of 270 km/s, similarly to the Sun.  

Once the calibration over the $m_\chi-\sigma_0$ grid is complete, we perform a single simulation at each point using the optimized nuisance parameters. This simulation yields the $r_{02}$ values for the model, which are then compared with observed values of $\chi^2_{r_{02}}$ to assess the goodness of the fit. 

\begin{figure}
        \begin{center}
            \includegraphics[width=1.0\linewidth]{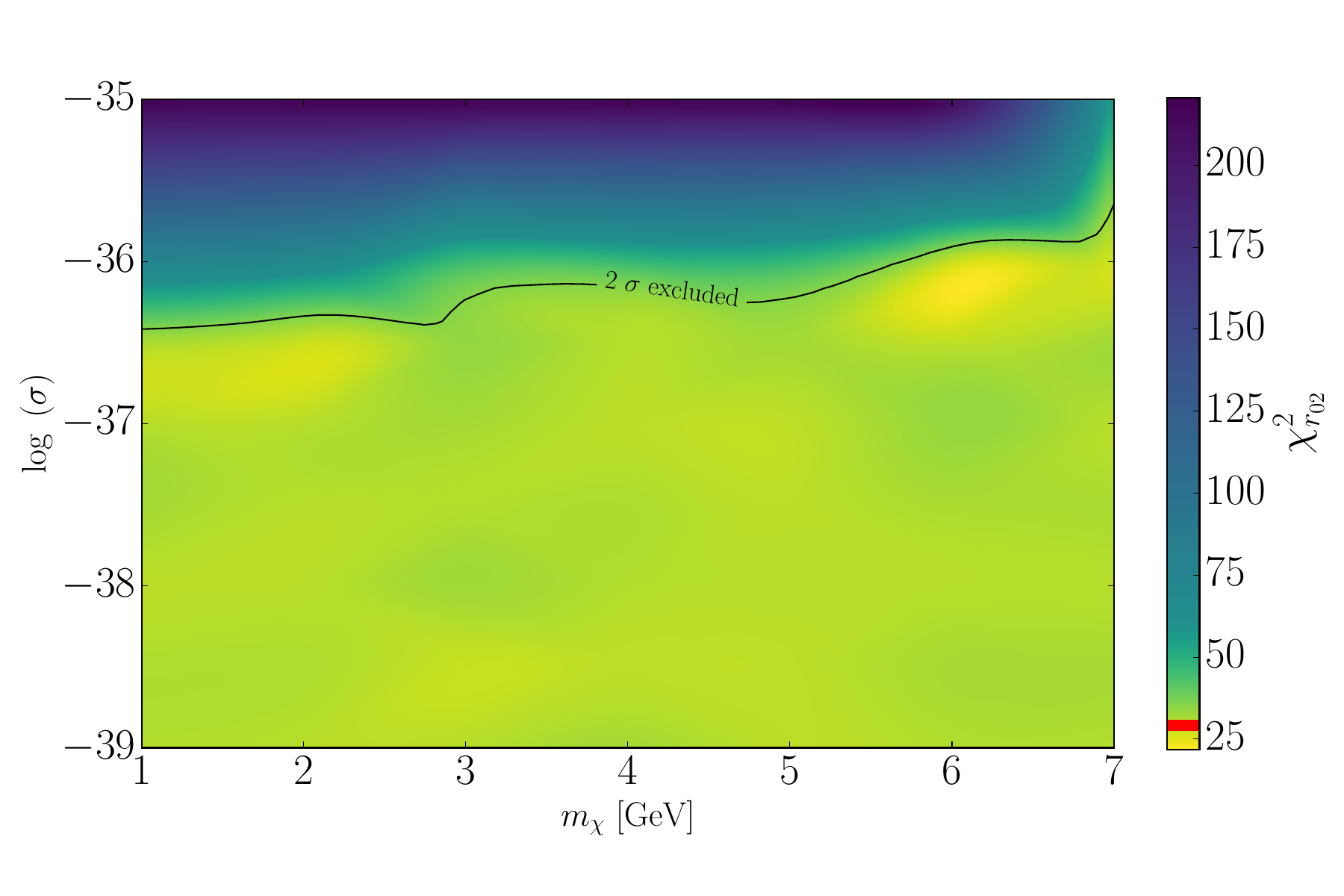}
            \caption[The $\chi^2_{r_{02}}$ values for a grid of $m_\chi-\sigma_0$ combinations for constant DM-nucleon interactions]{\textit{$\chi^2_{r_{02}}$ for a range of DM masses and cross sections} with constant DM-nucleon interactions in KIC 8228742. The solid black line represent the contours at which the modelled $\chi^2_{r_{02}}$ is $2\sigma$ worse than the no-DM case. The significance is estimated assuming the statistic follows a $\chi^2$ distribution with 2 degrees of freedom. The red line on the colourbar represents the no-DM $\chi^2_{r_{02}}$ value. } 
            \label{fig:contourPlot_hydrogen_r02}
        \end{center}
\end{figure}

\begin{figure}
    \centering
    \includegraphics[width=0.9\linewidth]{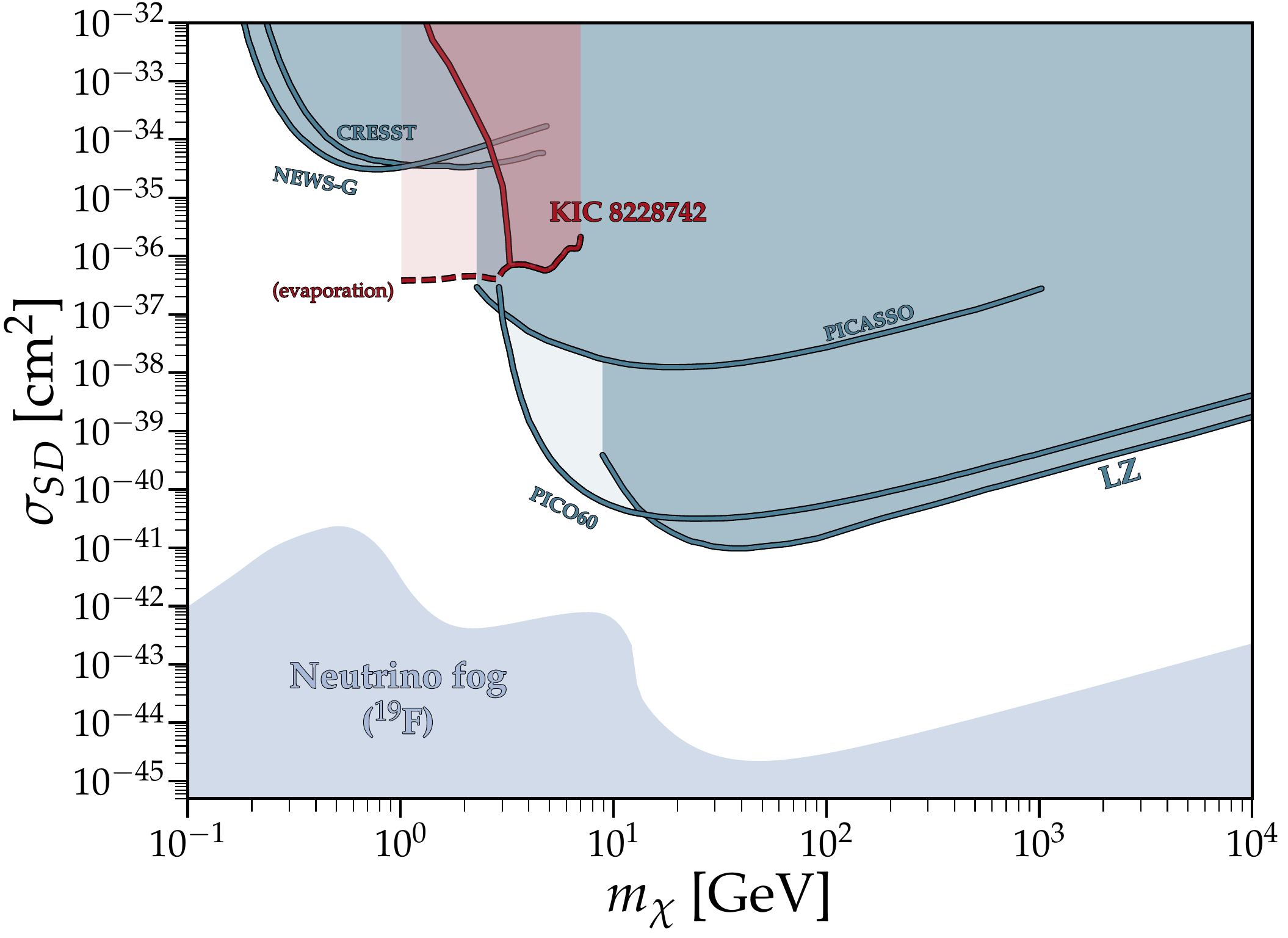}
    \caption{Excluded region (shaded red) based on asteroseismological data from the star KIC 8228742. The light region above the dotted line is likely not valid as it falls below the approximate evaporation mass for this star. Also shown are recent constraints from the CRESST \cite{CRESST:2019jnq}, NEWS-G \cite{NEWS-G:2017pxg}, PICASSO \cite{PICASSO:2012ngj}, PICO-60 \cite{PICO:2019vsc} and LZ \cite{LZ:2024zvo}, and the approximate location of the neutrino fog. Figure created with \href{https://github.com/cajohare/DirectDetectionPlots}{\texttt{DirectDetectionPlots}}.}
    \label{fig:DDexclusion}
\end{figure}

We begin by exploring the DM-nucleon scenario. Figure \ref{fig:contourPlot_hydrogen_r02} illustrates the $\chi^2_{r_{02}}$ for constant SD DM-nucleon interactions. The colour indicates the $\chi^2_{r_{02}}$ value for $m_\chi-\sigma_0$ combinations, while the no-DM case is represented by a red line on the colour bar. The solid contour line marks the region where the DM model performs $2 \sigma$ worse than the no-DM model. We place these results in context of direct detection experiments in Fig. \ref{fig:DDexclusion}. Taken at face value, the results presented in Fig. \ref{fig:contourPlot_hydrogen_r02} cover novel parameter space below $m_\chi = 3$ GeV, overtaking existing constraints by two orders of magnitude. However, in this region, evaporation is expected to become important, rapidly depleting the captured dark matter population. We approximate this region using solar values taken from Ref. \cite{raghuveer2017dark}. This region is shaded more lightly, and bounded by a dashed line in Fig. \ref{fig:DDexclusion}.

\begin{figure}
        \begin{center}
            \includegraphics[width=1.0\linewidth]{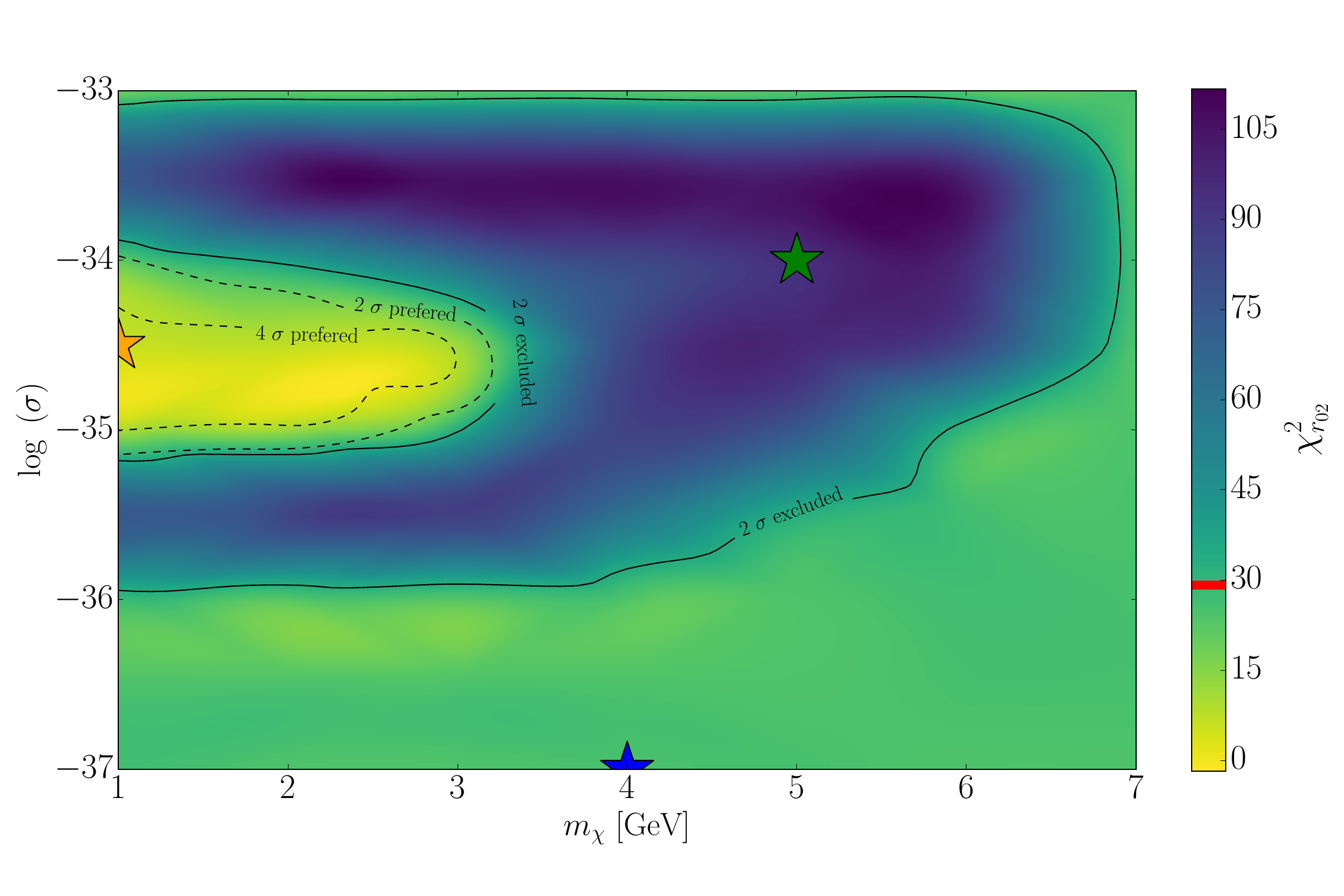}
            \caption[The $\chi^2_{r_{02}}$ values for a grid of $m_\chi-\sigma_0$ combinations for constant DM-electron interactions]{\textit{$\chi^2_{r_{02}}$ for a range of DM masses and cross sections} with constant DM-electron interactions in KIC 8228742. The solid black lines represent the contours at which the modelled $\chi^2_{r_{02}}$ is $2\sigma$ worse than the no-DM case. The dashed black lines show the $2\sigma$ and $4\sigma$ improvements in $\chi^2_{r_{02}}$ relative to the no-DM case. The significance is estimated assuming the statistic follows a $\chi^2$ distribution with 2 degrees of freedom. The red line on the colourbar represents the no-DM $\chi^2_{r_{02}}$ value. The green, orange, and blue stars represent Model A, B, and C given in Table \ref{tab:three_DM_models_electrons_r02}.}
            \label{fig:contour_r02_KIC_electrons}
        \end{center}
\end{figure}

Next, we turn to the DM-electron scenario. Figure \ref{fig:contour_r02_KIC_electrons} illustrates the $\chi^2_{r_{02}}$ for constant DM-electron interactions similarly to Figure \ref{fig:contourPlot_hydrogen_r02}. The solid contour lines denote regions where the DM model performs $2 \sigma$ worse than the no-DM model. The dashed contour lines mark regions where the DM model is $2\sigma$ and $4\sigma$ better than the no-DM model. 

Certain trends seen in the DM-nucleon scenario, which are due to the capture and transport formalisms, are expected to hold for DM-electron as well. For instance, DM is expected to have a significant impact on the star only within intermediate cross section ranges where $K \approx K_0$, meaning that heat transport is optimized. This is seen in Figure \ref{fig:contour_r02_KIC_electrons}, where $\sigma_0 \lesssim 10^{-36} \text{cm}^2$ and  $\sigma_0 \gtrsim 10^{-33} \text{cm}^2$ result in $\chi^2_{r_{02}}$ values that closely resemble the no-DM value.

Intermediate cross sections lead to a notable deviation in $\chi^2_{r_{02}}$ compared to the no DM case. 
Figure \ref{fig:contour_r02_KIC_electrons}, the solid black lines represent contours where the inclusion of DM resulted in $\chi^2_{r_{02}}$ values $2\sigma$ worse than that in the no-DM case. 
These combinations of $m_\chi-\sigma_0$ led to the reduction in the size of the convective core but did not entirely erase it. Conversely, the yellow-coloured region indicates an improvement to the fit of the frequency separation ratios relative to the no-DM case. These $m_\chi-\sigma_0$ combinations erase the convective core of the star. We observed improvements of up to $4\sigma$ in $\chi^2_{r_{02}}$, with the optimal point being a DM mass of 1 GeV and a cross section of $\sigma_0 \simeq  3 \times 10^{-34} \ \text{cm}^2$.

\begin{figure}
        \begin{center}
            \includegraphics[width=0.8\linewidth]{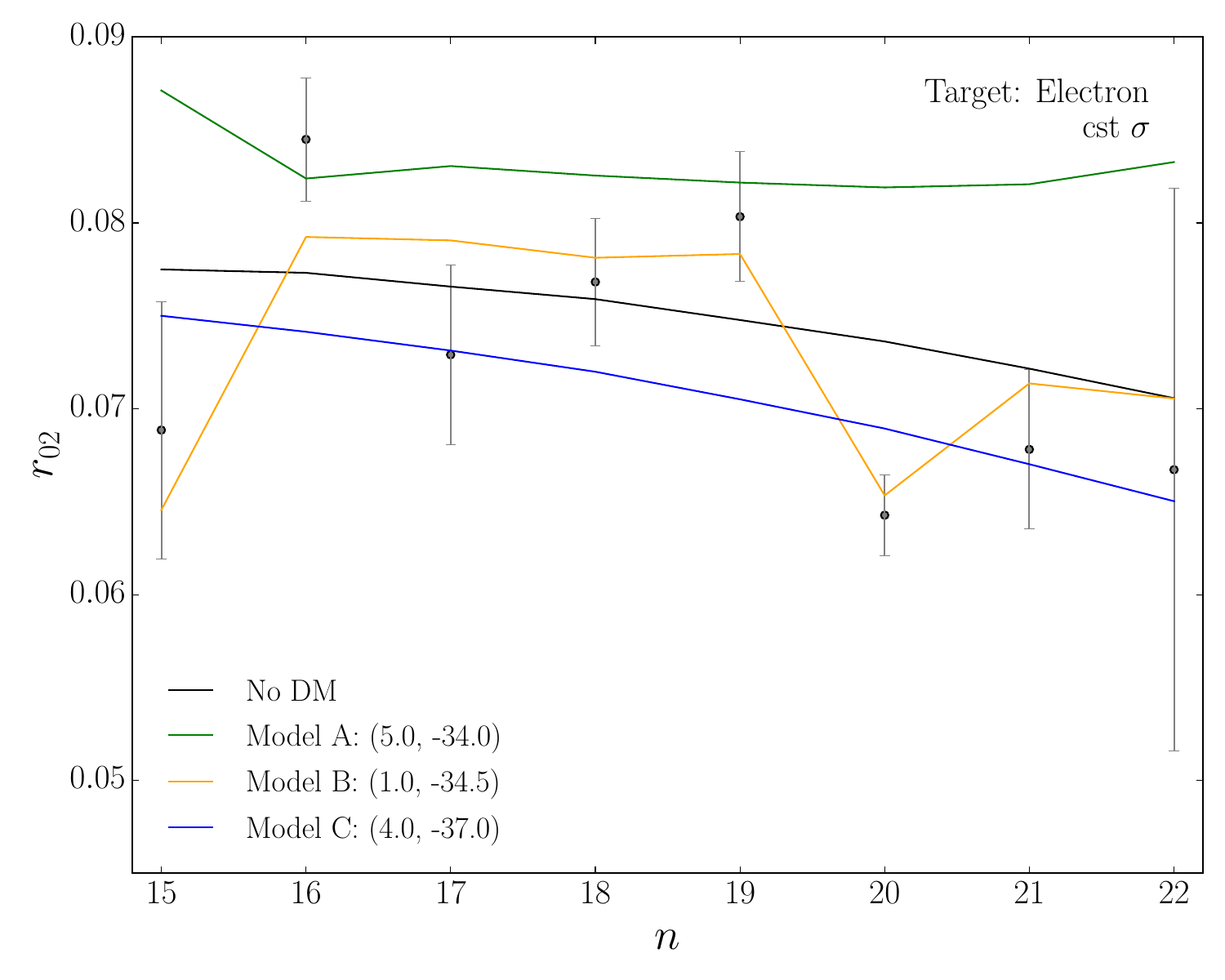}
            \caption[The frequency separation ratios for three models with constant DM-electron interactions.]{\textit{The frequency separation ratios} for the models in Table \ref{tab:three_DM_models_electrons_r02} (solid lines) and the observational ratios as calculated from the observed frequencies (black dots) \cite{appourchaux2012oscillation}. The values between parentheses specify ($m_\chi$, $\log (\sigma_0 /\mathrm{cm}^2)$) for each model.}
            \label{fig:r02_v_n_electrons_3_pts}
        \end{center}
\end{figure}

\begin{figure}
        \begin{center}
            \includegraphics[width=1.0\linewidth]{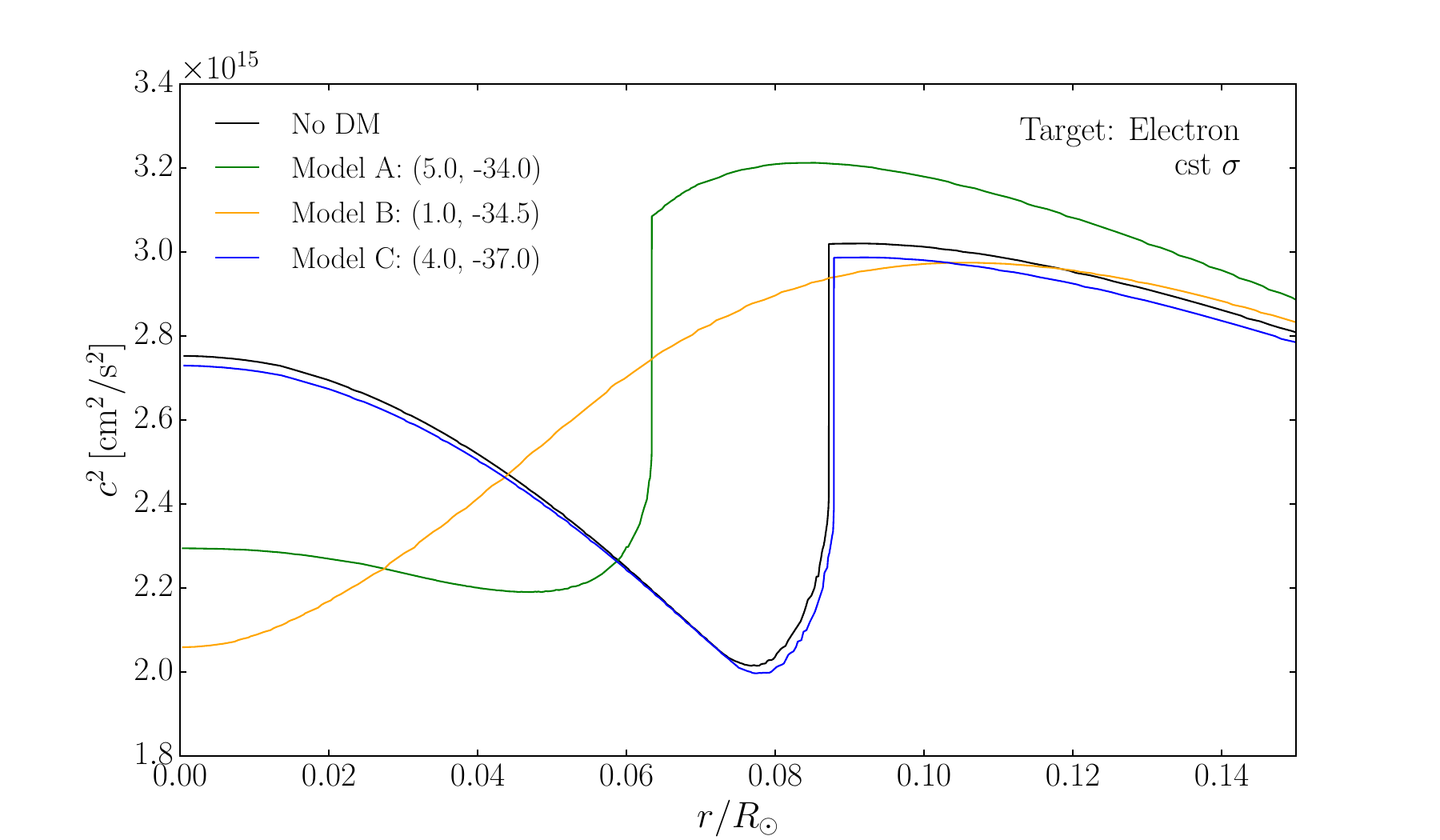}
            \caption[The sound speed profile for three models with constant DM-electron interactions.]{\textit{Squared sound speed as a function of radius} for the models in Table \ref{tab:three_DM_models_electrons_r02}. The values between parentheses specify ($m_\chi$, $\log \sigma_0$) for each model.}
            \label{fig:c2_v_radius_electrons_3_pts}
        \end{center}
\end{figure}

To further explore these effects, we plot the frequency separation frequency ratios for representative points labelled models A, B, and C in Figure \ref{fig:r02_v_n_electrons_3_pts}.
The DM mass and cross section for these points are given in Table \ref{tab:three_DM_models_electrons_r02} and are denoted by stars in Figure \ref{fig:contour_r02_KIC_electrons}. For the no-DM case and Model C (which is in the green region), the best fit for $r_{02}$ follows a linear trend. This fit is consistent with the findings from Figure 4 in \cite{Rato:2021tfc}. While the linear fits for the no-DM case and Model C are not identical, they result in similar stellar models, as indicated by the radial sound speed profiles $c^2(r)$, shown in Fig.~\ref{fig:c2_v_radius_electrons_3_pts}. The discrepancy arises because DM has a slight impact on the star due to its heat transport. For instance, Model C reduces the core temperature by 0.8 \% relative to the no-DM model. 
Model A, which shrinks the convective core, leads to a poorer fit and an overall overestimation of the $r_{02}$ values. Model B is the best fit model, where the $r_{02}$ values across radial orders $n$ follow the trend shown by the observational $r_{02}$ values. 

\begin{table}[h]

    \begin{tabular}{|c|c|c|c|c|c|c|c|c|}
        \hline
         Model & $m_\chi \ [\text{GeV}]$ & $\log \frac{\sigma_0}{\mathrm{cm}^2}$ & $T_{\text{eff}} \ [K]$ & $L \ [L_\odot]$ & $\log g$  & $\Delta\nu \ [\mu \text{Hz}]$ & $\chi^2_{\text{star}}$ & $\chi^2_{r_{02}}$ \\
        \hline
        Observ.  & $-$ & $-$ & 6042 $\pm$ 84 &  4.57 $\pm$ 1.45 & 3.92  $\pm$ 0.08 & 62.1 $\pm$ 0.13 & $-$ & $-$ \\
         No DM  & $-$ & $-$ & 6028.2 & 4.20 & 3.99 & 62.1 & 0.082 & 29.16\\
         Model A  & 5.00 & -34.0 & 6096.5 & 4.31 & 3.98  & 62.10 & 0.11 & 93.51\\
         Model B  & 1.00 & -34.5 & 5900.78 & 3.84 & 3.98 & 62.10 & 0.396 & 5.99\\
         Model C  & 4.00 & -37.0 & 6068.2 & 4.31 & 3.99  & 62.10 & 0.092 & 24.60 \\
        \hline
    \end{tabular}
    \caption[Table detailing the resulting observables and $\chi^2$ values for selected calibrated models.]{\textit{The models shown in Figure \ref{fig:r02_v_n_electrons_3_pts} and Figure \ref{fig:c2_v_radius_electrons_3_pts}.} The columns are: The DM mass, the cross section, the modelled effective temperature, the modelled effective luminosity, the log of the surface gravity, the large frequency separation, the $\chi^2_{\text{star}}$ form the initial calibration process (Equation \ref{equation:chi2Star}), and the $\chi^2_{r_{02}}$ for the modelled $r_{02}$.}
    \label{tab:three_DM_models_electrons_r02} 
\end{table}


A literal interpretation of the $\sim 4\sigma$ improvement over the standard model remains difficult, as the DM-electron cross sections required are many orders of magnitude larger than what is currently allowed by direct detection experiments, notably the XENON1T S2-only search \cite{aprile2019light}. Though relatively nearby, substructure in the DM distribution near KIC 8228742 could plausibly be different from the solar neighbourhood. However, reconciling these two results via a simple rescaling of the local DM density $\rho_0$ would imply that the density around KIC 8228742 was far higher than what galactic kinematics suggest the local density should be on Gyr time scales.

 Much of this region of DM masses also lies dangerously close to where evaporation is expected to be significant. Though some assumptions lead to an evaporation mass below a GeV \cite{raghuveer2017dark}, it is unclear how robust these are (though other mechanisms could potentially suppress evaporation, see e.g. \cite{Acevedo:2023owd}). Nonetheless, Fig. \ref{fig:contour_r02_KIC_electrons} clearly indicates a preferences for some non-standard extra luminosity inside a nearby star. The fact that this preference exists for DM-electron interactions, but not DM-nucleon couplings, may also provide a hint.

\section{Conclusions}
\label{sec:conclusion}
We have performed the first realistic stellar simulations of dark matter-nucleon and dark matter-electron interactions in a star that uses the newly-calibrated Spergel \& Press formalism, shown in Refs.~\cite{Banks:2021sba,Banks:2024eag} to be more accurate and robust than previous methods. We have focused on asteroseismological signatures in stars slightly heavier than the Sun, confirming that dark matter interactions can erase convective cores, leading to changes in various asteroseismological observables. 

In Sec.~\ref{sec:realstar}, we compared simulation predictions to a real star, KIC 
8228742, which should form a convective core in the absence of new physics. Using data from this single star, we obtained constraints on dark matter-nucleon interactions that are near the current strongest limit from direct detection experiments. Dark matter-electron interactions offer limits, but also a tantalizing 4$\sigma$ improvement over the standard model in some parts of the parameter space. 

As mentioned above, this parameter space is already excluded at face value, so a dark matter interpretation would require some work to reconcile these results with earth-based laboratory constraints. More precise data, and comparison with results in other nearby stars, will also help shed light on this intriguing result.

This initial study of constant DM-electron interactions on KIC 8228742 highlights it as a promising target for future more comprehensive investigations. For instance, we could perform a similar analysis as done in \cite{raen2021effects} by varying the DM density and stellar velocity dispersion, exploring different environments. Altering these parameters can impact our results: higher DM density increases capture rates, while lower relative velocities between the star and the infalling DM enhance capture probabilities. 
Also, investigating $v$-dependent and $q$-dependent interactions could yield better fits not constrained by DD limits.

Lastly, the observational constraints used for the calibration process have large uncertainties, limiting the effectiveness of the calibration. Having more accurate observational constraints can allow us to better calibrate the star and allow us to perform a better and more accurate analysis.

\acknowledgments
We thank Neal Avis Kozar for helpful input.  ACV is grateful to members of the LPPC (Harvard) and CTP (MIT) for their hospitality while completing this work. SB was supported by FRQNT and NSERC. ACV was supported by Arthur B. McDonald Canadian Astroparticle Physics Research Institute, NSERC the Canada Foundation for Innovation and the Province of Ontario. Research at Perimeter Institute is supported by the Government of Canada through the Department of Innovation, Science, and Economic Development, and by the Province of Ontario.

\bibliographystyle{JHEP_pat}
\bibliography{refs}

\appendix

\section{Non-constant cross sections}
\label{sec:appendix}

Interactions can be spin-dependent (SD) interactions which occur when DM couples to the nucleon spin, mediated by axial-vector or pseudoscalar particles; And spin-independent (SI) interactions which involve DM coupling coherently to the entire nucleus, with scalar or vector mediators \cite{goodman1985detectability}. 
The SD and SI DM-nucleus interaction cross sections are given by 
\begin{equation}
    \sigma^{SD}_{i, 0} =  \frac{m_{\text{nuc}}^2 (m_\chi +m_p)^2}{m_{\text{p}}^2 (m_\chi +m_{
\text{nuc}})^2} \ \frac{4 (J_i+1)}{3 J_i} | \langle S_{p,i} \rangle + \langle S_{n,i} \rangle |^2 \ \sigma_0 ,
\end{equation}
and 
\begin{equation}
    \sigma^{SI}_{i, 0} =  \frac{m_{\text{nuc}}^2 (m_\chi +m_p)^2}{m_{\text{p}}^2 (m_\chi +m_{
\text{nuc}})^2}A_i^2 \ \sigma_0,
\end{equation}
respectively, where $m_{\text{nuc}}$ is the nucleus mass, $m_p$ is the nucleon mass, $J_i$ is the total angular momentum of nucleus $i$, $\langle S_{p,i} \rangle$ and $\langle S_{n,i} \rangle$ are the expectation values of the spins of protons and neutrons averaged over all nucleons, and $A_i$ is the mass number. In this work we limit ourselves to SD interactions, therefore the coupling that is mainly probed is the one with protons
because almost all DM interactions are off hydrogen, and so the SD cross section is reduced to $\sigma^{SD}_{i, 0} = \sigma_0$.
The simplest models assume a DM-target differential cross section which is independent of the kinematics of the interaction and is simply given by a cross section $\sigma_0$.
However, the form of the interaction strength between DM and the target does not necessarily need to be constant and can depend on the relative velocity $v_{\text{rel}}$ and the transferred momentum $q$. These models are motivated by particle physics where often the interaction cross-section depends on the center of mass energy and the transferred momentum, which in the non-relativistic regime become the relative velocity and the transferred momentum. For instance, momentum-dependent interactions can arise from parity-violating couplings, small anapole or dipole interactions \cite{Sigurdson:2004zp, Chang:2009yt, Feldstein:2009tr, Chang:2010en}.

We define our DM-nucleon and DM-electron differential cross section as

\begin{equation}\label{eq:diff_sig_q}
\frac{d \sigma}{d \cos{\theta}} = \sigma_0 \left( \frac{q}{q_0} \right)^{2n_q},
\end{equation}
for momentum-dependent interactions, and

\begin{equation}\label{eq:diff_sig_v}
\frac{d \sigma}{d \cos{\theta}} = \sigma_0 \left( \frac{v_{\text{rel}}}{v_0} \right)^{2n_v},
\end{equation}

for velocity-dependent interactions, where $\theta$ is the center-of-mass scattering angle, $n_{v/q} = \{0, 1, 2 \}$, and $q_0$/$v_0$ are a reference momentum/velocity. We take the reference velocity to be the typical halo DM velocity, $v_0 = 220 \: \text{km s}^{-1}$, and the transfer momentum to be $ q_0 = 40 \: \text{MeV}$. 

This work specifically examines elastic scattering of DM with nucleons and electrons, focusing exclusively on spin-dependent interactions. Such interactions primarily occur with hydrogen for nucleons, thus eliminating the need to consider nuclear form factors required for heavier nucleons.

\subsection{Capture}
\label{sec:qvcapture}

The expressions for the capture rate can be generalized to non-constant interactions.
 For $v/q$-dependent interactions, the differential scattering rate is given by 

\begin{equation}
\begin{split}
 R_{T, n_q, n_v}^- (w \rightarrow v) = \frac{2^{5+2n_v+n_q}}{\sqrt{\pi}} \frac{\mu_+^{4+2(n_v+n_q)}}{u_T^3(r)} \frac{v}{w} n_T(r) \frac{m_\chi^{2n_q} \sigma_0}{q_0^{2 n_q} v_0^{2 n_v}} \\ C_{n_q, n_v} \left[ F_{n_q, n_v}(w-v) e^{- \alpha^2_-} - G_{n_q, n_v}(w+v) e^{-\alpha_+^2} \right. \\ \left. + H_{n_q,n_v} \chi(-\alpha_-, \alpha_+) + J_{n_q, n_v} \chi(-\beta_-, \beta_+) e^{\mu(w^2-v^2)/u_T^2(r)} \right], 
\end{split}
\end{equation}
where $C_{n_q, n_v}$, $F_{n_q, n_v}$, $G_{n_q, n_q}$, $H_{n_q, n_v}$ and $J_{n_q, n_v}$ are interaction dependent. For $v^2$ and $q^2$, they are given in Appendix A of \cite{raghuveer2017dark}. For $v^4$, and $q^4$, the $C_{n_q, n_v}$ factors are given in Table \ref{tab:RMinus_funcs}, and $F_{n_q, n_v}$, $G_{n_q, n_q}$, $H_{n_q, n_v}$ and $J_{n_q, n_v}$ were derived in our work and are given by
\begin{equation}
\begin{split}
F_{2, 0} = 2(8+20\mu+7\mu^2)u_e^2 + (8\mu+7\mu^2+4\mu^3+\mu^4)v^2 \\
+ 2(5\mu^2+4\mu^3+\mu^4)v w +(8\mu+7\mu^2+4\mu^3+\mu^4)w^2,
\end{split}
\end{equation}
\begin{equation}
\begin{split}
G_{2, 0} = 2(8+20\mu+7\mu^2)u_e^2 + (8\mu+7\mu^2+4\mu^3+\mu^4)v^2 \\
- 2(5\mu^2+4\mu^3+\mu^4)v w +(8\mu+7\mu^2+4\mu^3+\mu^4)w^2,
\end{split}
\end{equation}

\begin{equation}
H_{2, 0} = (8 + 5 \mu (4 + 3 \mu)) u_e^4 + 4 \mu (2 + 5 \mu) u_e^2 w^2 + 4 \mu^2 w^4,
\end{equation}
\begin{equation}
J_{2, 0} = (8 + 5 \mu (4 + 3 \mu)) u_e^4 + 4 \mu (2 + 5 \mu) u_e^2 v^2 + 4 \mu^2 v^4,
\end{equation}
and
\begin{equation}
F_{0,2} = 6u_e^2+\mu^2(v + w)^2,
\end{equation}
\begin{equation}
G_{0,2} = 6u_e^2+\mu^2(v - w)^2,
\end{equation}
\begin{equation}
H_{0,2} = 12 u_e^4 + \mu^2(v^2-w^2)^2+6\mu u_e^2(w^2-v^2),
\end{equation}
\begin{equation}
J_{0,2} = 12 u_e^4 + \mu^2(v^2-w^2)^2+6\mu u_e^2(v^2-w^2).
\end{equation}

\begin{table}[h]
    \centering
    \renewcommand{\arraystretch}{2.0}
    \begin{tabular}{|c|c|c|c|c|c|c|}
        \hline
         $n_v$ & $n_q$ & $C_{n_q, n_v}$ \\
        \hline
        1 & 0 &  $\frac{u_T^5}{2^5 \mu_+^4 \mu }$  \\
        2 & 0 & $\frac{u_T^3}{2^{10} \mu^2 \mu_+^6}$   \\
        0 & 1 & $\frac{\pi u_T^5}{4 \mu_+ \mu^2}$  \\
        0 & 2 & $\frac{\pi u_T^3}{2^3 \mu^4 \mu_+^2}$  \\
        \hline
    \end{tabular}
    \caption{Analytical forms for $C_{n_q, n_v}$ appearing in the differential scattering rate for the different interactions.}
    \label{tab:RMinus_funcs}
\end{table}

 The thermal average of the cross section is 
\begin{equation}
    \langle \sigma (\boldsymbol{v}) \rangle = \int_0^{\infty} \left( \int_{-1}^{1} \frac{d \sigma}{d \ \cos \theta} d \cos \theta \right)  f(v_{\text{rel}})d v_{\mathrm{rel}},
\end{equation}

where $\frac{d \sigma}{d \ \cos \theta}$ is the differential cross section (Equation \ref{eq:diff_sig_q} and Equation \ref{eq:diff_sig_v}), $v_{\text{rel}}$ is the relative velocity between DM and the target, and $ f(v_{\text{rel}})$ is the normalized distribution for the relative velocity given by 
\begin{equation}
     f(v_{\text{rel}}) = \frac{ v_{\text{rel}}^2}{[\pi v_T^2(1+\mu)]^{3/2}} e^{-v_{\text{rel}}^2/v_{T}^2(1+\mu)}.
\end{equation}
The resulting thermally averaged cross section  is given in Table \ref{tab:thermal_average_sigma} for constant and momentum/velocity dependent interactions. 

\begin{table}[h]\label{table:thermallyAveragedSigma}
    \centering
    \renewcommand{\arraystretch}{1.7} 
    \begin{tabular}{|c|c|c|}
        \hline
         $n_v$ & $n_q$ & $\langle \sigma (\boldsymbol{v}) \rangle$ \\
        \hline
        0 & 0 & $2\sigma_0$ \\
        1 & 0 & $3 (1+\mu) \frac{u_T^2}{v_0^2} \sigma_0$ \\
        2 & 0 & $\frac{15}{2} (1+\mu)^2 \frac{u_T^4}{v_0^4} \sigma_0$ \\
        0 & 1 &  $6 \frac{m_\chi^2}{(1+\mu)} \frac{u_T^2}{q_0^2} \sigma_0$ \\
        0 & 2 & $40 \frac{m_\chi^4}{(1+\mu)^2} \frac{u_T^4}{q_0^4} \sigma_0$ \\
        \hline
    \end{tabular}
    \caption{Thermally averaged cross section for the different interactions}
    \label{tab:thermal_average_sigma}
\end{table}

\subsection{Heat transport}
\label{sec:heattransportderivation}
The approximate energy transport was first derived by Spergel \& Press \cite{Spergel:1984re}, and generalized to non-constant cross section in Ref. \cite{Banks:2021sba}. Here we will lay out the general steps for this integration here given that the original calculation does not include all the steps and has several typos.

We want to find the radial energy transport profile for interactions between the DM and a target population where both populations follow Maxwellian distribution functions given by 
\begin{equation}
f_{\chi} = \left( \frac{m_\chi}{2 \pi T_\chi} \right)^{3/2} n_{\chi} \exp \left( \frac{-m_\chi v_\chi^2}{2 \:  T_\chi} \right),
\end{equation}
and
\begin{equation}
f_{T} = \left( \frac{m_T}{2 \pi T_T (r)} \right)^{3/2} n_{T} \exp \left( \frac{-m_T v_T^2}{2 \: T_T(r)} \right),
\end{equation}
respectively. The energy transferred in a single collision is given by 

\begin{equation}
\Delta E  = \frac{m_\chi}{2} \left( v_{f,\chi}^2 - v_{i, \chi}^2 \right), 
\end{equation}
where $v_{f, \chi}$ and $v_{i, \chi}$ are the final and initial DM velocities, respectively. We take the average over the center of mass scattering angle to get the average energy transfer per collision 

\begin{equation}
\langle \Delta E \rangle = \frac{m_\chi \: m_T}{(m_\chi + m_T)^2} (1 - Q) \left[ m_T v_T^2 - m_\chi v_\chi^2 + (m_\chi - m_T) v_\chi v_T \cos \theta \right], 
\end{equation}
where $\theta$ is the scattering angle between the incoming DM and target particles and $Q$ is given by 

\begin{equation}
Q = \frac{\int \frac{d \sigma}{d \cos \theta_{cm}} \cos \theta_{cm} d \cos \theta_{cm}}{\int \frac{d \sigma}{d \cos \theta_{cm}} d \cos \theta_{cm}}, 
\end{equation}
where $\theta_{cm}$ is the center of mass scattering angle. We can now evaluate the radial energy transfer profile per unit time per unit mass, $\epsilon$, using 

\begin{equation}
\epsilon(r) = \frac{1}{\rho (r)} \int d^3 v_\chi f_\chi \int d^3 v_T f_T \: \sigma_{\text{tot}} \langle \Delta E \rangle |\boldsymbol{v_\chi} - \boldsymbol{v_T}|. 
\end{equation}

We work in spherical coordinates and evaluate the trivial angular integrals to get 

\begin{equation}
\begin{split}
\epsilon = \frac{8}{\pi} \left( \beta_\chi \beta_T \right)^{3/2}\frac{m_\chi \: m_T}{(m_\chi + m_T)^2} (1-Q) \int_0^\infty v_\chi^2 dv_\chi e^{-\beta_\chi v_\chi^2} \int_0^\infty v_T^2 dv_T e^{-\beta_T v_T^2} \\
\times \int_{-1}^1 d \cos \theta ( m_T v_T^2 - m_\chi v_\chi^2 + (m_{\chi} - m_T) v_\chi v_T \cos \theta ) |\boldsymbol{v_\chi} - \boldsymbol{v_T}|, 
    \end{split}
\end{equation}

where $\beta_{i} = m_i/(2T_i)$ is the square of the thermal velocity.

To evaluate the magnitude of the velocity difference, we need to split the energy transfer expression into two terms: a term with $v_\chi > v_T$ and a term with $v_\chi \leq v_T$. We can denote these two terms by $I_{v_\chi > v_T}$ and $I_{v_\chi \leq v_T}$ and so 

\begin{equation}
\epsilon = I_{v_\chi > v_T} + I_{v_\chi \leq v_T},
\end{equation}

and they are given by 
\begin{equation}
\begin{split}
I_{v_\chi > v_T} = \frac{8}{\pi} \left( \beta_\chi \beta_T \right)^{3/2}\frac{m_\chi \: m_T}{(m_\chi + m_T)^2} (1-Q) \int_0^\infty v_\chi^2 dv_\chi e^{-\beta_\chi v_\chi^2} \int_0^{v_\chi} v_T^2 dv_T e^{-\beta_T v_T^2} \\
\times \int_{-1}^1 d \cos \theta ( m_T v_T^2 - m_\chi v_\chi^2 + (m_{\chi} - m_T) v_\chi v_T \cos \theta ) (v_\chi^2 + v_T^2 - 2 \: | \boldsymbol{v_\chi}| \: | \boldsymbol{v_T}| \cos \theta),
    \end{split}
\end{equation}

and

\begin{equation}
\begin{split}
I_{v_\chi > v_T} = - \frac{8}{\pi} \left( \beta_\chi \beta_T \right)^{3/2}\frac{m_\chi \: m_T}{(m_\chi + m_T)^2} (1-Q) \int_0^\infty v_\chi^2 dv_\chi e^{-\beta_\chi v_\chi^2} \int_{v_\chi}^\infty v_T^2 dv_T e^{-\beta_T v_T^2} \\
\times \int_{-1}^1 d \cos \theta ( m_T v_T^2 - m_\chi v_\chi^2 + (m_{\chi} - m_T) v_\chi v_T \cos \theta ) (v_\chi^2 + v_T^2 - 2 \: | \boldsymbol{v_\chi}| \: | \boldsymbol{v_T}| \cos \theta),
    \end{split}
\end{equation}
respectively. We perform a change in variables and introduce the variable $z^2 = v_\chi^2 + v_T^2 - 2 \: | \boldsymbol{v_\chi}| \: | \boldsymbol{v_T}| \cos \theta$ and so $zdz = -| \boldsymbol{v_\chi}| \: | \boldsymbol{v_T}| \:  d \cos \theta$. We can now rewrite our integrals as 
\begin{equation}
\begin{split}
I_{v_\chi > v_T} = \frac{8}{\pi} \left( \beta_\chi \beta_T \right)^{3/2}\frac{m_\chi \: m_T}{(m_\chi + m_T)^2} (1-Q) \int_0^\infty v_\chi^2 dv_\chi e^{-\beta_\chi v_\chi^2} \int_0^{v_\chi} v_T^2 dv_T e^{-\beta_T v_T^2} \\
\times \int_{v_\chi - v_T}^{v_\chi + v_T} \left[ m_T v_T^2 - m_\chi v_\chi^2 + \frac{1}{2}(m_{\chi} - m_T) (v_\chi^2 + v_T^2 - z^2) \right]\frac{z^2 dz }{v_\chi v_T},
    \end{split}
\end{equation}

and

\begin{equation}
\begin{split}
I_{v_\chi > v_T} = \frac{8}{\pi} \left( \beta_\chi \beta_T \right)^{3/2}\frac{m_\chi \: m_T}{(m_\chi + m_T)^2} (1-Q) \int_0^\infty v_\chi^2 dv_\chi e^{-\beta_\chi v_\chi^2} \int_0^{v_\chi} v_T^2 dv_T e^{-\beta_T v_T^2} \\
\times \int_{v_T - v_\chi}^{v_\chi + v_T} \left[ m_T v_T^2 - m_\chi v_\chi^2 + \frac{1}{2}(m_{\chi} - m_T) (v_\chi^2 + v_T^2 - z^2) \right]\frac{z^2 dz }{v_\chi v_T}.
    \end{split}
\end{equation}

These integrals can be evaluated to get the analytic expression for the radial energy transport 
\begin{equation}\label{equation:energyTransportFinal}
\epsilon(r) = \frac{A_{n_q, n_v}}{\rho(r)}\sqrt{\frac{2}{\pi}} \frac{m_\chi m_T}{(m_\chi+m_T)^2} n_\chi (r) n_T (r) (1-Q) \sigma_{\text{tot}} \left( T_\chi - T(r) \right) \left( \frac{T(r)}{m_T} + \frac{T_\chi}{m_\chi} \right)^{\frac{1}{2}+n},
\end{equation}

where $n = 0, 1, 2$ for constant, $v^2/q^2$, and $v^4/q^4$ interactions, respectively, and where $A_{n_q, n_v}$ and $(1-Q) \sigma_{\text{tot}}$ are interaction dependent and given in Table \ref{tab:A_Q_Factors}.
\begin{table}[h]
    \centering
    \renewcommand{\arraystretch}{1.7} 
    \begin{tabular}{|c|c|c|c|}
        \hline
         $n_v$ & $n_q$ & $A_{n_q, n_v}$ & $(1-Q) \sigma_{\text{tot}}$ \\
        \hline
        0 & 0 & 8 & $2 \sigma_0$ \\
        1 & 0 & 48 & $2 \sigma_0/v_0^{2}$  \\
        2 & 0 & 348 & $2 \sigma_0/v_0^{4}$ \\
        0 & 1 & 48 & $\frac{16}{3} \frac{m_\chi^2 \sigma_0}{(1+\mu)^2 q_0^2}$ \\
        0 & 2 & 348 &  $16 \frac{m_\chi^4 \sigma_0}{(1+\mu)^4 q_0^4}$ \\
        \hline
    \end{tabular}
    \caption[The interaction dependent factors needed for calculating the transported energy.]{The interaction dependent $A_{n_q, n_v}$ and $(1-Q) \sigma_{\text{tot}}$ factors which appear in Equation \ref{equation:energyTransportFinal}}
    \label{tab:A_Q_Factors}
\end{table}

\begin{figure}
    \centering
    \includegraphics[width=0.45\linewidth]{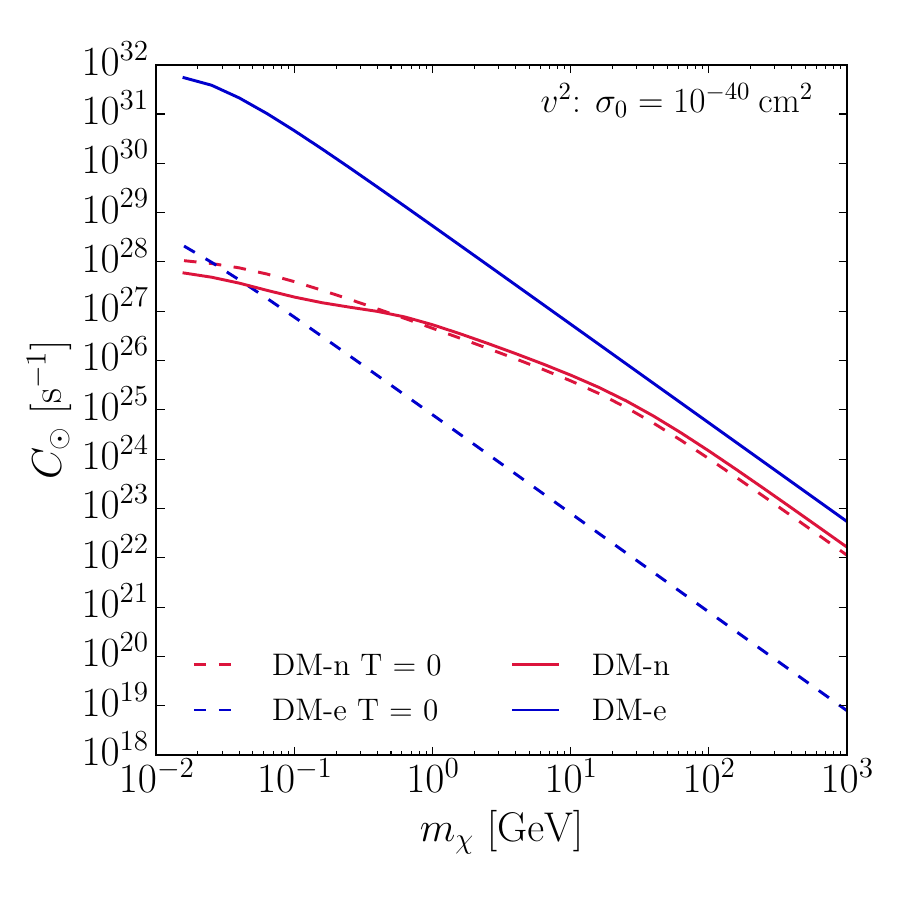}\includegraphics[width=0.45\linewidth]{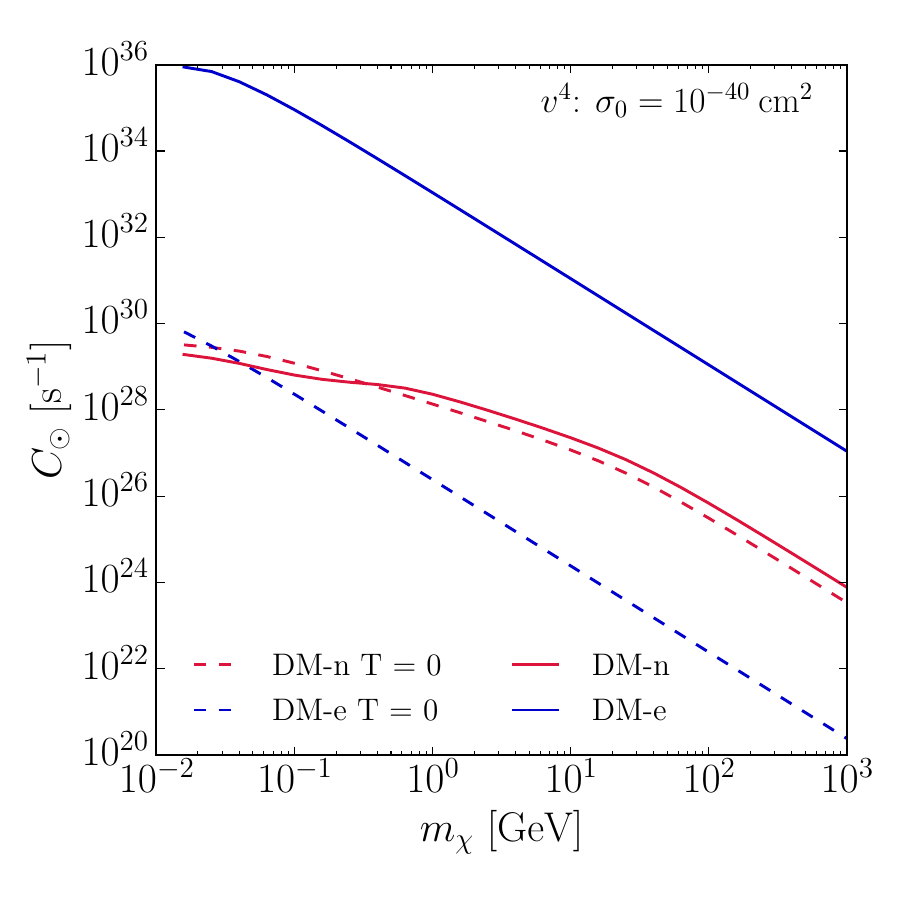} \\
    \includegraphics[width=0.45\linewidth]{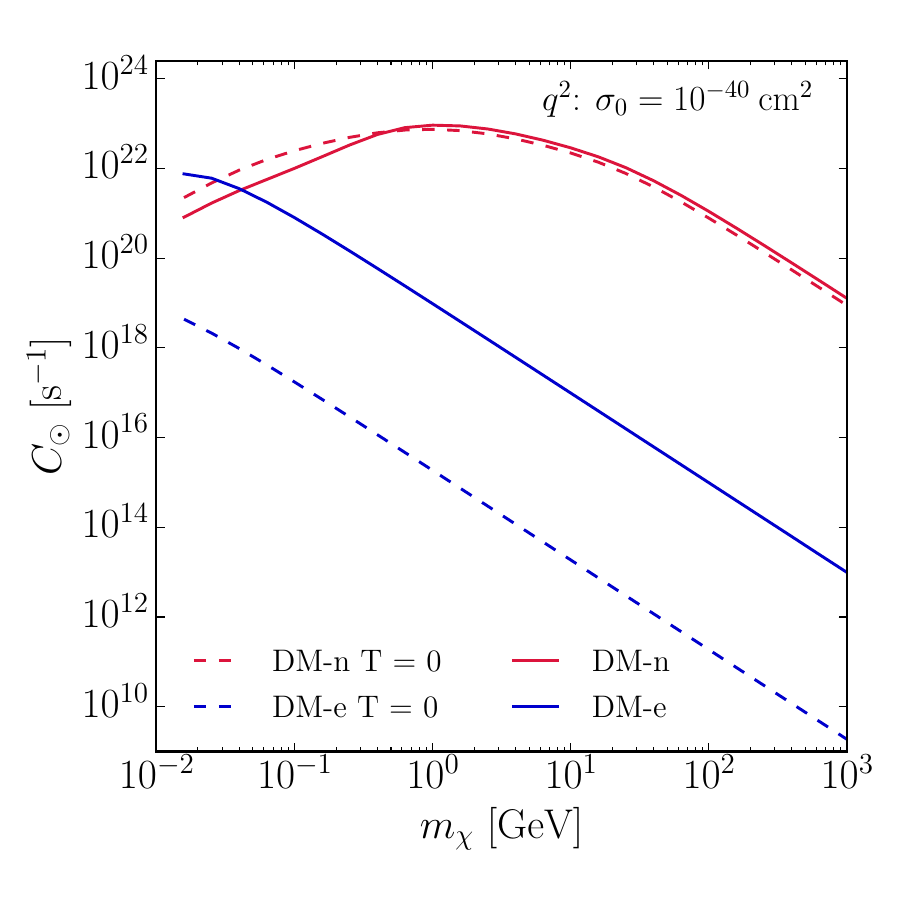}
    \includegraphics[width=0.45\linewidth]{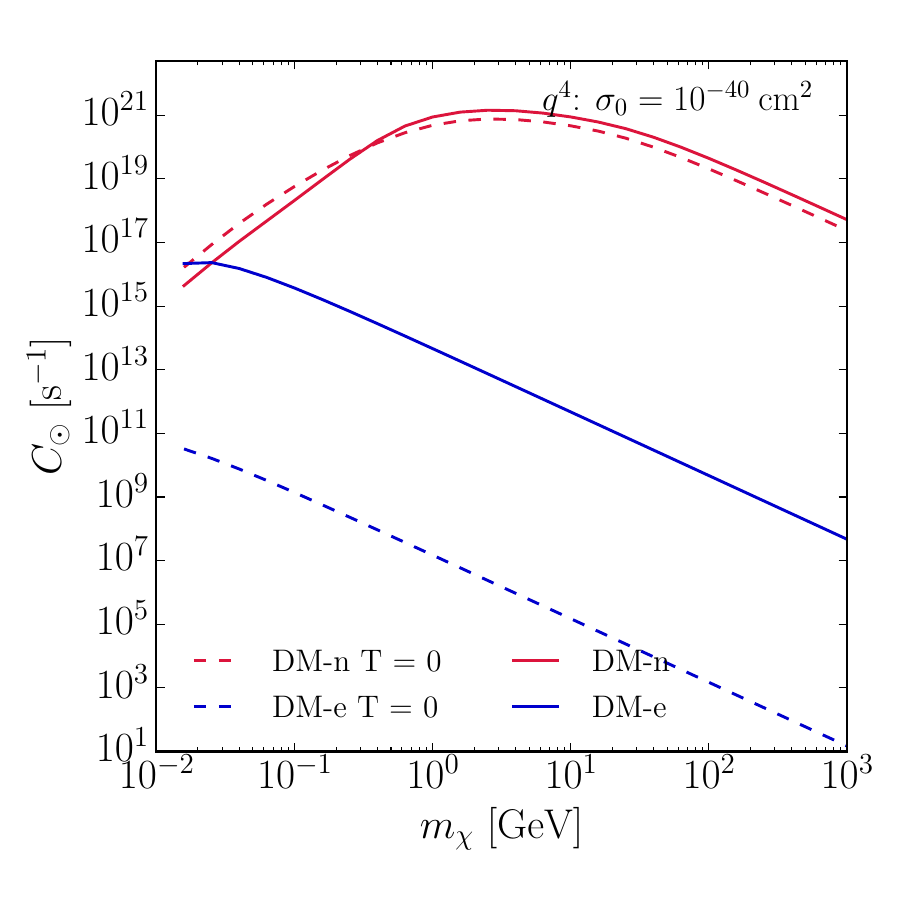}
    \caption{Dark matter capture rate in the Sun for non-constant DM-nucleon (red) and DM-electron (blue) interactions. Line styles show the result of including (solid) or neglecting (dashed) the thermal velocity of the target particles. }
    \label{fig:enter-label}
\end{figure}

\begin{figure}
    \centering
    \includegraphics[width=0.45\linewidth]{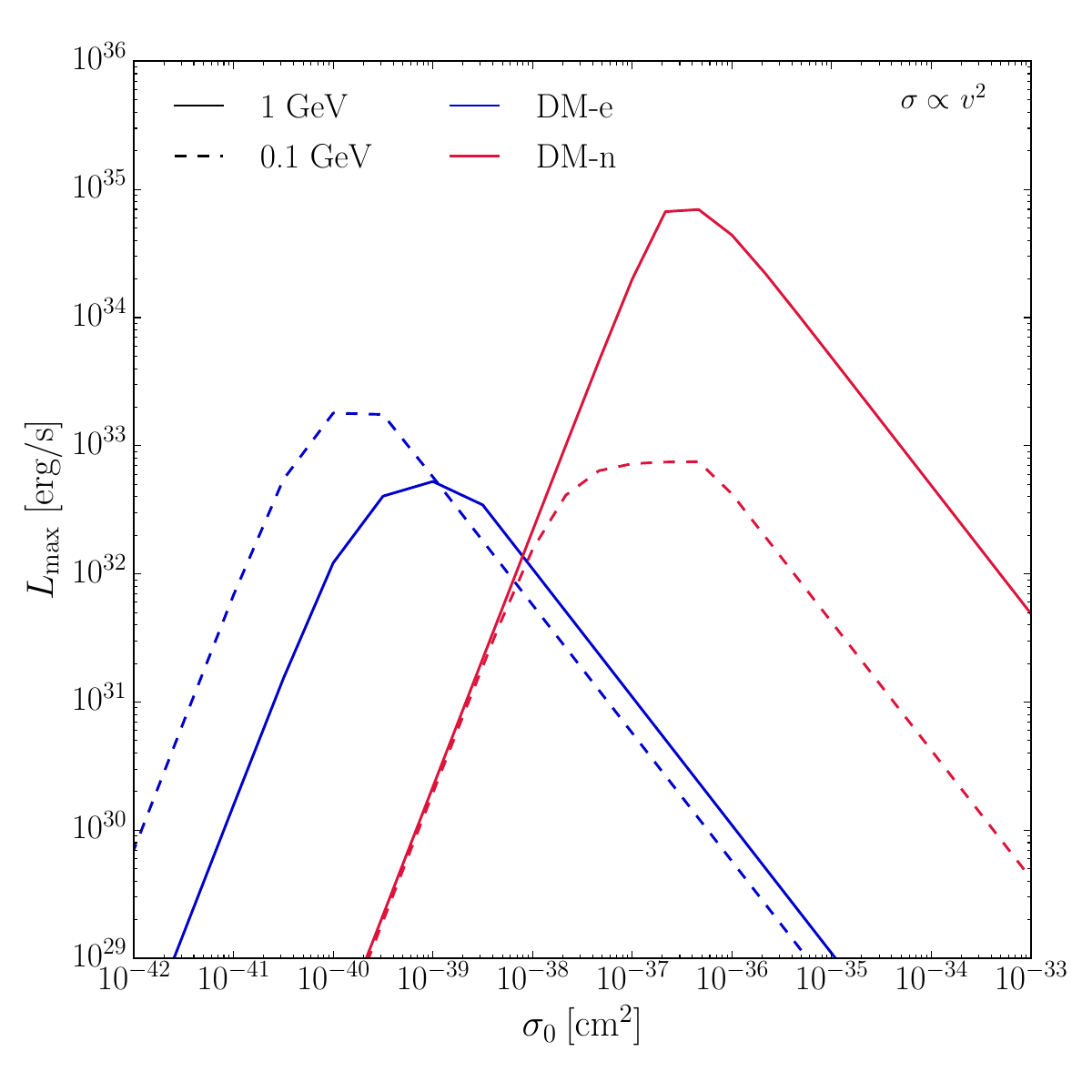}\includegraphics[width=0.44\linewidth]{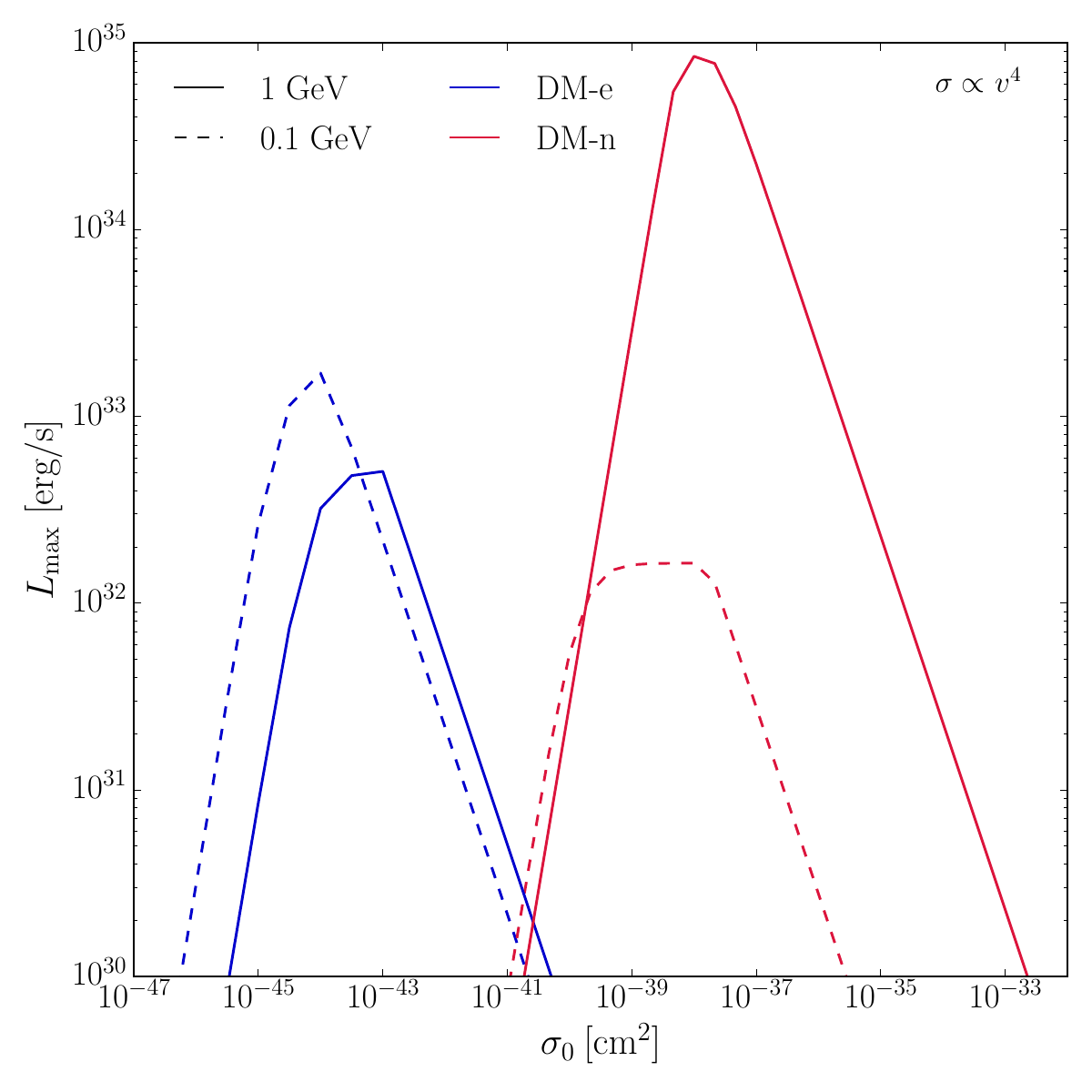}\\
    \includegraphics[width=0.45\linewidth]{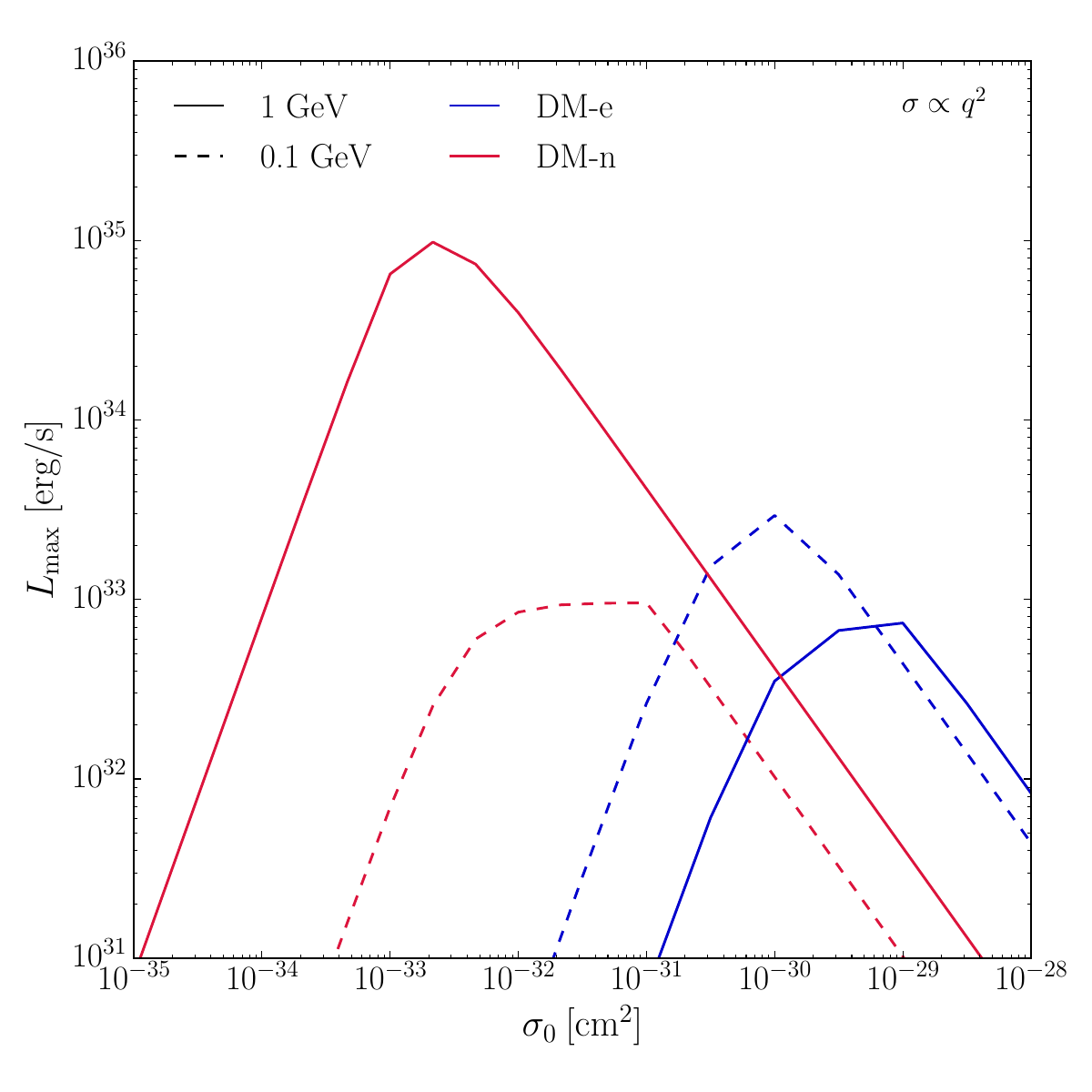} \includegraphics[width=0.45\linewidth]{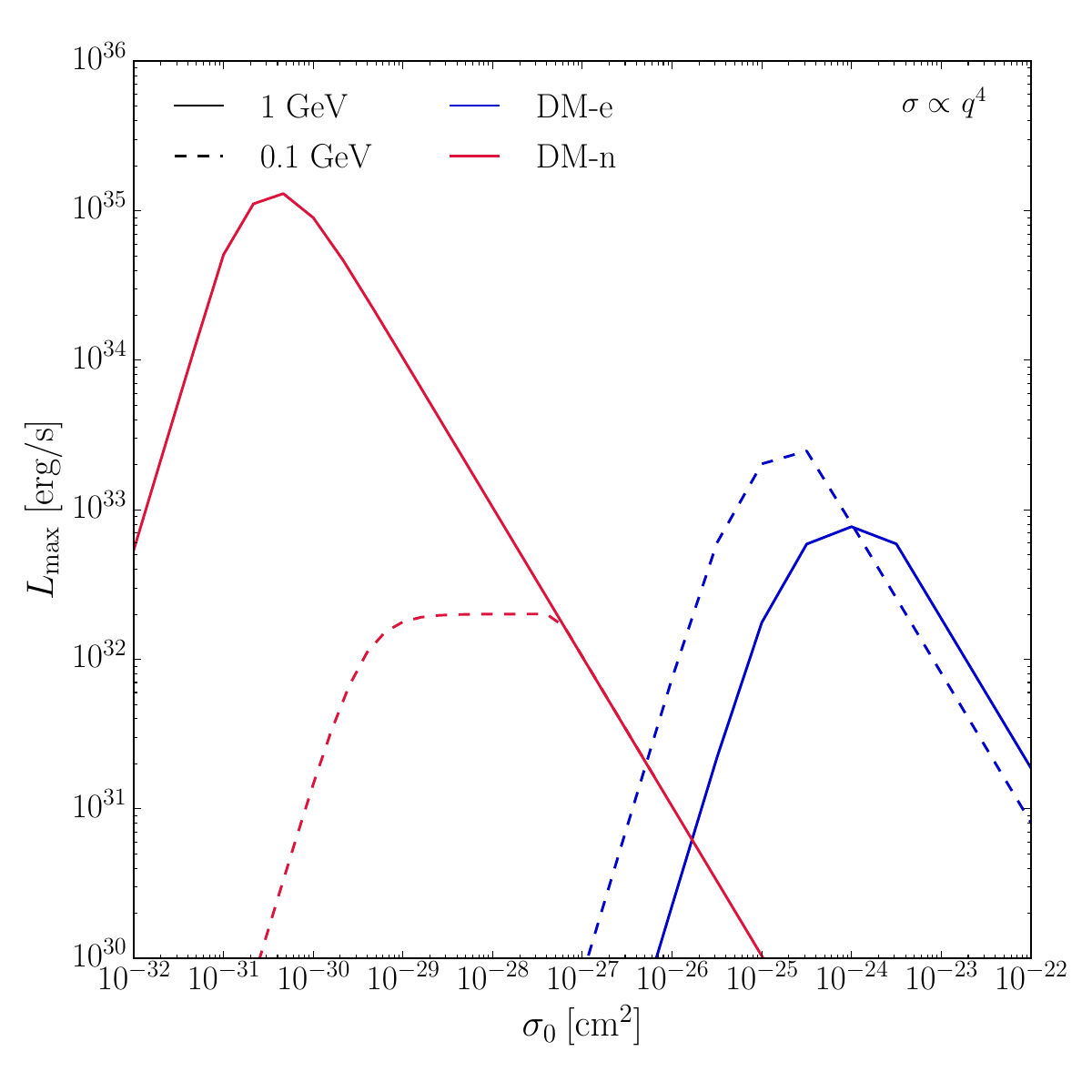}
    \caption{Heat transported as a function of cross section, for velocity and momentum-dependent cross sections as defined in Eqs. (\ref{eq:diff_sig_q},\ref{eq:diff_sig_v}). }
    \label{fig:lumvnqn}
\end{figure}


\end{document}